\documentclass[sigconf]{acmart}

\AtBeginDocument{%
  }

\setcopyright{acmlicensed}

\copyrightyear{2026}
\acmYear{2026}
\setcopyright{cc}
\setcctype{by-nc-nd}
\acmConference[CHI '26]{Proceedings of the 2026 CHI Conference on Human Factors in Computing Systems}{April 13--17, 2026}{Barcelona, Spain}
\acmBooktitle{Proceedings of the 2026 CHI Conference on Human Factors in Computing Systems (CHI '26), April 13--17, 2026, Barcelona, Spain}
\acmPrice{}
\acmDOI{10.1145/3772318.3791572}
\acmISBN{979-8-4007-2278-3/2026/04}

\usepackage{multirow, booktabs}
\usepackage{algorithm}
\usepackage{algorithmic}
\usepackage{xcolor}
\definecolor{promptfg}{HTML}{432A63} 
\newcommand{\prompt}[1]{\textit{\textcolor{promptfg}{#1}}}

\begin{document}


\title[VidTune: Creating Video Soundtracks with Generative Music and Contextual Thumbnails]{VidTune: Creating Video Soundtracks with\\Generative Music and Contextual Thumbnails}


\author{Mina Huh}
\affiliation{
  \institution{University of California, Berkeley}
  \country{Berkeley, California, USA}}
\email{minahuh@berkeley.edu}

\author{C. Ailie Fraser}
\affiliation{
  \institution{Adobe Research}
  \country{Seattle, Washington, USA}}
\email{fraser@adobe.com}

\author{Dingzeyu Li}
\affiliation{
  \institution{Adobe Research}
  \country{Seattle, Washington, USA}}
\email{dinli@adobe.com}

\author{Mira Dontcheva}
\affiliation{
  \institution{Adobe Research}
  \country{Seattle, Washington, USA}}
\email{mirad@adobe.com}

\author{Bryan Wang}
\affiliation{
  \institution{Adobe Research}
  \country{Seattle, Washington, USA}}
\email{bryanw@adobe.com}



\begin{abstract}
\revised{Music shapes the tone of videos, yet creators find it hard to find soundtracks that match their video's mood and narrative.}
Recent text-to-music models let creators generate music from text prompts, but our formative study (N=8) shows creators struggle to construct diverse prompts, quickly review and compare tracks, and understand their impact on the video. We present VidTune, a system that supports soundtrack creation by generating diverse music options from a creator’s prompt and producing contextual thumbnails for rapid review. VidTune extracts representative video subjects to ground thumbnails in context, maps each track’s valence and energy onto visual cues like color and brightness, and depicts prominent genres and instruments. Creators can refine tracks with natural language edits, which VidTune expands into new generations. 
\revised{In a controlled user study (N=12) and an exploratory case study (N=6), participants found VidTune helpful for efficiently reviewing and comparing music options and described the process as playful and enriching.}

\end{abstract}

\begin{teaserfigure}
  \centering
  \vspace{-10pt}
  \includegraphics[width=\textwidth]{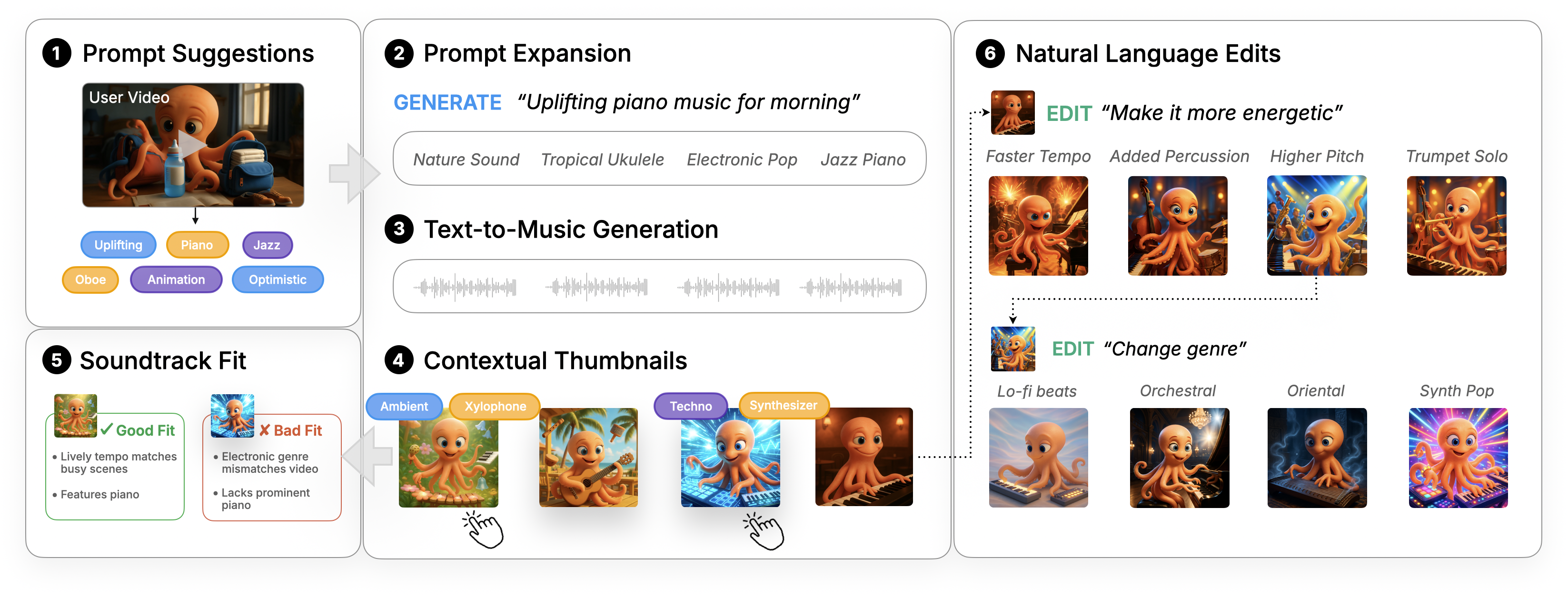}
  \vspace{-20pt}
  \caption{VidTune is an interactive system that helps video creators generate soundtracks. VidTune provides prompt suggestions based on the input video (1) and expands the user's prompt (2) to generate a diverse set of candidate music tracks (3). Each track is presented with \textit{contextual thumbnails} (4) for efficient music preview in-context, and users can hover over each track to see reusable keywords and a \textit{soundtrack fit check} (5). Creators can iteratively refine tracks with language-guided edits (6), which VidTune turns into updated candidate tracks.} 
  \Description{Three-panel VidTune workflow (left→right): (1) Prompt Suggestions—frame from a cartoon octopus video with tags Uplifting, Piano, Jazz, Oboe, Animation, Optimistic; (2) Prompt Expansion—Generate “Uplifting piano music for morning” with suggested terms Nature Sound, Tropical Ukulele, Electronic Pop, Jazz Piano; (3) Text-to-Music Generation—audio waveform; (4) Contextual Thumbnails—four thumbnails (Ambient, Xylophone, Techno, Synthesizer) with a hand cursor selecting; (5) Soundtrack Fit—Good Fit (lively tempo, features piano) vs Bad Fit (electronic genre mismatches video, lacks prominent piano); (6) Natural-Language Edits—“Make it more energetic” → Faster Tempo, Added Percussion, Higher Pitch, Trumpet Solo; “Change genre” → Lo-fi beats, Orchestral, “Oriental,” Synth Pop; solid arrows show the main flow (1→5), dotted arrows show edit branches from thumbnails.}
  \label{fig:teaser}
\end{teaserfigure}

\newcommand{\ipstart}[1]{\vspace{1mm}\noindent{\textbf{\textit{#1.}}}}
\newcommand\smallverb[1]{\texttt{\small #1}}

\newcommand{\revised}[1]{\textcolor{black}{#1}}

\maketitle
\section{Introduction}
\revised{Generative music is increasingly used in video production~\cite{perez_youtube_2025, hammad2025s}, letting creators generate custom soundtracks from natural language prompts. However, current text-to-music workflows often struggle to meet the demands of creating video soundtracks. Creators may lack a clear sense of what music would fit their video, and even when they do, they often struggle to articulate these intentions as effective prompts~\cite{hammad2025s, yakura2023iteratta}.
Our formative study with 8 video creators surfaced these same breakdowns in practice, with participants relying on only a few familiar descriptors. 
Moreover, as these models rapidly produce many alternatives, reviewing, organizing, and aligning tracks with the video’s structure became a major bottleneck~\cite{fu2025exploring}. 
The temporal nature of audio forced them to listen through each track to compare and recall favorites, making it hard to skim and triage options. 
These issues were even more pronounced for deaf and hard-of-hearing (DHH) creators, who lacked accessible cues to explore and validate the music.}



\revised{One way to mitigate these challenges is to represent audio in visual forms that reveal similarities and differences at a glance, in ways that are difficult to achieve through listening alone.
Motivated by this, 
we present VidTune, a text-to-music generation tool that uses \textit{contextual thumbnails} to visually summarize generative music outputs in the context of the user's video. In VidTune, each thumbnail takes the form of both a static image and a short animated video generated from it. These thumbnails map music attributes such as mood, energy, and instrumentation onto visual style and composition with subjects anchored in the user's video, giving each track a concise and meaningful visual summary. This explicit audio-visual mapping helps address gaps in current tools, which rely on vague titles, generic cover art, and waveforms that may not be intuitive for many creators, including non-music experts and those with limited hearing. }

\revised{
Grounded in prior work and our formative insights, we designed VidTune around four capabilities: helping creators \textbf{explore} by expanding user prompts into musically diverse alternatives, \textbf{review} with contextual thumbnails that visually distinguish different tracks, \textbf{refine} by steering generations with natural language edits, and \textbf{manage} large sets of outputs via an overview of the music space that reveals explored directions and potential next steps.
}

Our technical evaluation shows that VidTune’s prompt expansion algorithm can increase musical diversity and that its thumbnails more reliably reflect the generated music than a baseline. 
We also evaluated VidTune in a within-subjects study with 12 video creators, who compared VidTune against a baseline interface similar to existing AI music generation tools. Participants described VidTune as more expressive and enjoyable to use, and felt the results were more worth the effort. 
Finally, an exploratory case study (N=6) using creators’ own videos further shows how the thumbnails were perceived in situ and highlights VidTune’s potential to make music generation more accessible to a broader set of users.


Taken together, our work makes the following contributions:
\begin{itemize}
    \item VidTune, a system that enables creators to explore, interpret, and refine generative music for video soundtracks.
    \item An AI pipeline that generates \textit{contextual thumbnails} by capturing key elements of the input video and the generated music tracks to support \textit{visual sensemaking} of music.  
    
    \item Empirical findings from a controlled study (N=12) demonstrating VidTune’s advantages over a baseline, and an exploratory study (N=6) highlighting emergent uses and broader applications for contextual thumbnails.
\end{itemize}




\section{Related Work}
VidTune is informed by prior work on video-based music generation, music visualization, and sensemaking of generative AI output.

\subsection{Music Generation for Videos}
Recent advances in AI assist creators in composing melodies~\cite{ding2024songcomposer}, harmonizing chords~\cite{lim2017chord, kim2025amuse}, and even writing lyrics~\cite{suno, ding2024songcomposer}. 
Modern text-to-music models~\cite{cideron2024musicrl, suno, udio} also make end-to-end music generation more accessible to novices~\cite {louie2020novice} and creators with hearing impairments~\cite{choi2025exploring} \revised{by allowing them to describe desired moods or styles in natural language.} 
\revised{
Other work conditions audio directly on video content: synthesizing \textit{diegetic audio} aligned with on-screen actions~\cite{gan2020foley, su2020audeo, du2023conditional}, and composing \textit{background music} to match narrative emotion~\cite{rubin2012underscore, rubin2014generating, di2021video, xie2025filmcomposer}, or visual rhythm~\cite{li2024muvi, tian2025vidmuse, li2024vidmusician}.
More recently, commercial text-to-video models (\textit{e.g.,} Sora~\cite{openai_sora_2025}, Veo~\cite{google_deepmind_veo3_2025}) can produce both visuals and synchronized audio from a single prompt, effectively generating end-to-end videos with soundtracks. However, these systems keep music generation opaque and non-interactive for creators.}

Beyond optimization-centric approaches, recent HCI systems help video creators easily express their music goals by suggesting text prompts from video analysis~\cite{hammad2025s}, allowing users to provide example songs as inspiration~\cite{frid2020music}, \revised{or iteratively updating music based on a conversational dialogue with the user~\cite{zhang2023loop}}. These systems highlight the importance of supporting high-level user input, previewing music in the context of the user's video, and balancing automated support with user control.
However, they mainly focus on initial prompting and refinement of a few outputs, with less attention to how creators can efficiently understand, compare, and evaluate the large number of options produced in music creation workflows~\cite{fu2025exploring, huang2020ai}. We address this gap by designing workflows for in-context review and comparison of generative soundtracks. 

\subsection{Music Visualization}\label{sec:rw_musicviz}
Music is inherently temporal and non-visual, which makes it harder to skim or grasp at a glance compared to images or video. Visualizations can help by externalizing key attributes to augment the listening experience~\cite{liu2023generative}, evoke synesthetic effects~\cite{lee2025musicolors}, and improve accessibility~\cite{yi2025toward}. \revised{Symbolic and signal-level encodings such as scores and waveforms are commonly used for composition, analysis, and performance~\cite{mattias2023sound, lima2021survey}, and video editing and stock music platforms often rely on waveform displays to help users navigate and edit music. While useful for precise editing, they do not convey perceptual qualities of music such as mood or vibe. In contrast, semantic visualizations emphasize mood, style, or structure to aid listening and exploration~\cite{lima2021survey, khulusi2020survey} by mapping rhythm, timbre, and affect onto colors, motion, scenes, and art styles~\cite{davis2018visual,tendulkar2020feel,liu2023generative, chen2023musicjam, izzati2025musiscene}. 
Semantic visualizations can also aid music \textit{creation} by generating real-time images to inspire MIDI-based composition~\cite{yang2024exploring}, or in commercial tools like Suno~\cite{suno} and Udio~\cite{udio} by adding thumbnail images to generated songs.
However, these approaches are primarily designed to enrich the experience of creating a single track, rather than to help creators select from multiple generated soundtracks. We investigate what video creators seek when choosing a generative soundtrack and propose semantic visualizations that embed musical attributes into imagery anchored in the user's own video. 
Building on accessible music visualization work that maps musical mood and structure onto colors and shapes~\cite{deja2020vitune, wang2023music, mchugh2021towards}, or expressive faces~\cite{wang2023music}, we also explore how VidTune’s thumbnails can function as a non-auditory accessibility channel in selecting generative soundtracks.} 

Even with semantic visual aids, exploring and understanding a large set of music options can be overwhelming, especially as generative models make it possible to generate infinitely many candidates. Most music search and recommendation interfaces show results as ranked lists~\cite{chen2024background, epidemicsoundsite}, and music generation platforms similarly display outputs as lists of tracks ordered by recency~\cite{suno, udio}, but a linear layout does not convey relationships or organization across the space of choices~\cite{petridis2022tastepaths, gajdusek2021spotifygraph}. 
\revised{Prior work has used 2D maps to visualize music spaces for exploring large collections of songs~\cite{pampalk2002content, van2004mapping, julia2009songexplorer, takahashi2018instrudive} and visualizing users' taste ~\cite{petridis2022tastepaths, gajdusek2021spotifygraph}. Building on these approaches, we introduce a music map adapted for generative soundtrack selection, where the tracks generated for a given video are represented as thumbnails in a 2D space that enables exploration, comparison, and iteration.}

\subsection{Sensemaking of Generative AI Outputs in Creative Work}

Generative AI enables creators to explore design spaces more efficiently by streamlining creative processes~\cite{wang2024lave, huh2025videodiff} and generating diverse variations, including ones beyond their own capabilities~\cite{chung2021intersection}. \revised{In creative workflows, prior work shows that seeing alternatives side by side helps people notice differences~\cite{huh2025videodiff}, reflect on trade-offs, and achieve higher-quality outcomes~\cite{dow2010parallel}. Prior HCI systems operationalize this idea by expanding users' input prompts to generate more diverse results~\cite{cai2023designaid, han2025poet, brade2023promptify}.}

\revised{With many alternatives, interfaces must help users make sense of them. 
Recent sensemaking tools organize generations into hierarchies \cite{suh2024luminate, almeda2024prompting} and 2D similarity maps \cite{brade2023promptify}, and guide navigation with constraints and filters that help users traverse large collections~\cite{jeon2021fashionq, matejka2018dream, matejka2014video}.
Some focus on comparison, supporting users in understanding differences across outputs~\cite{huh2023genassist, benharrak2025historypalette, huh2025videodiff}: for instance, GenAssist describes similarities and differences across generated images~\cite{huh2023genassist}, and VideoDiff visualizes alternative video edits side by side~\cite{huh2025videodiff}.}


\revised{These systems focused on visual and textual domains, where alternatives can be quickly skimmed and compared with thumbnails, storyboards, or text summaries~\cite{woodruff2001using, bederson2001photomesa, goldman2006schematic, barua2025lotus}. 
Recent work on generative music has begun to explore constructing diverse prompts and iteratively refining tracks~\cite{hammad2025s, yakura2023iteratta}, but these interfaces typically present results as waveforms~\cite{hammad2025s} or spectrograms~\cite{yakura2023iteratta} that are difficult for non-experts to interpret and still require sequential listening. VidTune addresses this gap by presenting soundtrack options as contextual thumbnails that encode key musical attributes and providing organization mechanisms to support rapid comparison and navigation without listening to every track.}
\section{Formative Study}
To understand the strategies and challenges of using AI to generate music for videos, we conducted a formative study with 8 video creators. The formative study consisted of semi-structured interviews and a music generation task using participants' own videos. 
Unlike prior work that surveyed creators~\cite{frid2020music} or observed existing workflows without text-to-music tools~\cite{hammad2025s}, we observed creators scoring their own videos with a state-of-the-art text-to-music model to surface strategies and challenges that arise in actual use.





\subsection{Method}
\ipstart{Participants}
We recruited 8 video creators who add music to their videos on a regular basis (P1-P8, \S{A} Table~\ref{tab:participants}). Participants were recruited via mailing lists and compensated 40 USD for a 1.5-hour remote study conducted via Zoom. 
Participants had an average of 15 years of video creation experience (SD=7.43) and created a wide variety of videos, including short films, commercials, vlogs, and \revised{social-media reels.
We intentionally sampled a range of experience levels to capture how both less-experienced and skilled creators approach AI music tools.} 3 participants regularly used music generation tools including Suno~\cite{suno} and Udio~\cite{udio} (P1, P2, P6), 1 participant occasionally used them (P4), and 4 participants had never used them (P3, P5, P7-P8). We included 3 creators (P5, P7-P8) who were deaf or hard-of-hearing (DHH) as they encounter distinct perceptual challenges in generating music and can inform the design of an inclusive tool.

\ipstart{Procedure}
We first conducted a semi-structured interview asking participants how they currently find or generate music for their videos. 
\revised{Then, participants completed a 30-minute task where they used Suno~\cite{suno} to generate music for their own videos which ranged from 2 to 17 minutes in length.}
While there are many music generation tools available in the market today, we selected Suno as it is one of the most widely used text-to-music platforms with over 25 million users~\cite{musicbusinessworldwide}. For participants who had not used Suno, we provided a short tutorial on how to write prompts, configure generation settings, and export music. Participants could generate unlimited tracks and add multiple pieces to different parts of the video.
With DHH creators, we adopted their preferred communication channels (Zoom chat, live captions) and encouraged them to ask questions about each generated music to make decisions, which researchers answered in real time. We analyzed the questions to understand what information DHH creators seek in music generation, thereby guiding what attributes future systems should surface~\cite{huh2023genassist}.

We transcribed the interviews and participants’ comments during the task and grouped their strategies and challenges into 4 stages of the music-generation workflow: 1) writing prompts, 2) reviewing music, 3) iterating on outputs, and 4) organizing them.

\begin{table*}[htbp]
\centering
\sffamily
\renewcommand{\arraystretch}{1.15}    
\setlength{\tabcolsep}{0.6em}         

\begin{tabular}{p{0.18\textwidth} p{0.18\textwidth} p{0.64\textwidth}}
\toprule
\textbf{Category} & \textbf{Criteria} & \textbf{Example Question} \\
\midrule
\multirow{2}{*}{\parbox{0.18\textwidth}{Contextual Qualities}}
 & Prompt alignment & \textit{Does this music align with my text prompt/desired style?} \\
 & Video alignment  & \textit{How well does this music fit the visuals and mood of my video scene?} \\
\midrule
\multirow{6}{*}{\parbox{0.18\textwidth}{Musical Qualities}}
 & Genre \& Style           & \textit{What is the genre or musical style of this track?} \\
 & Mood \& Emotion          & \textit{What mood or emotion does this music convey?} \\
 & Energy \& Tempo          & \textit{What is the music’s overall energy level and tempo like?} \\
 & Instruments              & \textit{What instruments are prominently featured in this music?} \\
 & Structure \& Progression & \textit{How does the music's structure and progression unfold over time?} \\
 & Errors \& Quality        & \textit{Are there generation artifacts or quality issues?} \\
\bottomrule
\end{tabular}

\vspace{6pt}
\caption{Types of information that our formative study participants considered when evaluating generative music for their video soundtracks. Items were coded from participants’ explanations of how they evaluated tracks and, for DHH creators, from the questions they asked about each track.}
\label{tab:review_criteria}
\Description{Categories, criteria, and example questions for judging AI-generated music for video—Contextual Qualities: Prompt alignment (“Does this music align with my text prompt/desired style?”), Video alignment (“How well does this music fit the visuals and mood of my video scene?”); Musical Qualities: Genre \& Style (“What is the genre or musical style?”), Mood \& Emotion (“What mood or emotion does this music convey?”), Energy \& Tempo (“What is the overall energy level and tempo?”), Instruments (“What instruments are prominently featured?”), Structure \& Progression (“How does the structure/progression unfold over time?”), Errors \& Quality (“Are there generation artifacts or quality issues?”); items coded from participant explanations and, for DHH creators, from their questions.
}
\end{table*}

\subsection{Findings}
\ipstart{Current practices are tedious}
All participants described music as essential to shaping the story and audience experience. P8 noted \textit{``Most of my audience is hearing and I want them to also enjoy my videos.''} Similarly P3 stated \textit{``[Soundtracks] are very important, but with long videos, it's harder to come up with multiple music tracks.''}

To source soundtracks for their videos, 5 participants searched stock libraries (P2-4, P6, P8), 2 composed music themselves (P6-P7), and 4 used generative music tools (P1-P2, P4, P6). 
Among DHH creators, P8 often selected from recommended music on short-form platforms. To review whether the music is right for the video context, she checked lyrics, examined where the track had been used, and read audience comments about how the music felt. The other two DHH creators often delegated soundtrack selection to friends and family (P5, P7).

Despite these strategies, participants found it tedious to search for a track that suited their goals, which made generative music particularly appealing. As P6 explained, \textit{``When I have much time, I’d make my own music using FL Studio, but it’s hard now because I create and upload more often. Stock tracks all sound too similar; I don’t want to use a song people recognize from other channels.''} P1, who frequently uses AI music, highlighted its increasing quality: \textit{``Now people don’t know it’s AI made. It doesn’t sound like AI.''}







\ipstart{Coming up with diverse prompts is hard}
Creators with musical backgrounds (P1-P2, P4) used more concrete, musically specific prompts (\textit{e.g.,} genre, instrumentation, tempo, progression cues), while those with less musical background (P3, P5-P8) wrote more abstract or usage-oriented prompts tied to the video content and desired audience reaction. 3 participants wanted to tailor style to their audience but were unsure what fits. P6 noted, \textit{``This is a huge challenge for me... I am 46 years old; what I generate might feel outdated to students. What do they listen these days?''} 
Because starting from a blank prompt was challenging, participants drew on references and recommendations -- browsing YouTube examples (P3) and asking ChatGPT for ideas (P2, P6, P8).
2 participants also tried including artist names (P2—\textit{``Miley Cyrus''}; P1—\textit{``Hans Zimmer''}) as a concise way to convey style, but found these blocked by Suno due to copyright restrictions.
P4 expressed a desire to branch further: \textit{``I really want to see more different music styles but don't know how to get there.''} P1 noted that long prompts or explicit timing cues were often not reliably followed~\cite{li2024dance} and wanted ways to filter for generations that adhere to requests in the prompt.


These observations echo Hammad et al.~\cite{hammad2025s} on the need for video-grounded prompt suggestions. Further, our findings highlight the importance of helping users expand and explore broader options to discover what fits.

\ipstart{Reviewing generated music is time-consuming}
Table~\ref{tab:review_criteria} summarizes what creators consider when reviewing and selecting soundtracks. 
Beyond low-level musical qualities, participants evaluated whether a track adhered to the prompt and how well it suited the footage.
To evaluate music in context with their video, participants listened to music with a side-by-side still frame of the video (P3-P4, P6-P7) or imported tracks into a video editor to preview (P1-P2).
The Suno version used in our study did not allow users to specify duration and produced 2–4-minute tracks. Novices tended to listen longer (\textit{``I need 40–90 seconds to know if I like it,''} -- P3), whereas others skipped around to quickly understand the entire song.

Participants emphasized listening beyond a quick skim. P2 explained, \textit{``As songs are usually much longer than my scene, I need to listen through the whole piece to find the section that fits best.''}
P6 also noted \textit{``I need to verify the music holistically. But humans' instinct is to quickly judge, so that's hard. Similar to how AI-generated images have deformations when you look closely, generated music can be the same but harder to notice as it's audio. [...] Once, I heard female vocals gradually turn into to male later in the track and that was weird.''}
    

With the ability to quickly generate many options, participants described \textit{reviewing} as the bottleneck. The default titles and thumbnails provided by the platform were not informative enough to convey the music or distinguish tracks, so participants resorted to sequential listening. P3 remarked, \textit{``Thumbnails aren’t catchy or descriptive at all, and all titles are ``Untitled'' unless I manually type in. Just showing the prompt isn’t enough.''} Consequently, participants forgot how previous tracks sounded or which they preferred, often replaying tracks multiple times (P6 repeating \textit{``Was it this one?''}) to find them again.
Both P4 and P6 wanted visual snippets to preview music, and P6 added \textit{``I don’t want it [future system] to just choose the best music for my video [...] Overly suggesting can harm our creativity. I just want help in reviewing.''}

DHH creators encountered additional hurdles in reviewing generated music. Without familiar social or context cues (\textit{e.g.,} artist name, comments, prior uses), it was harder to evaluate generated tracks: \textit{``Does it actually sound like reggae music [as in the prompt]? [...] there’s no artist or comments, so I cannot really trust this music''} (P8). 
Both P5 and P7 wanted more visual support like colors or shapes to quickly convey emotion and intensity, as dense text descriptions can be slower to read. P5 stated \textit{``We deaf people love visual stuff. [...] I like music visualizers that convey emotion and intensity. But I can also see abstract visuals being more confusing when I'm selecting music not just enjoying it.''}

\ipstart{Iterating is necessary but non-trivial}
Participants were rarely satisfied with the first pass and went through multiple iterations -- either refining the prompt or using Suno’s audio edit feature -- to align generation with their music goal.
P2 wanted more diverse options for edits, explaining \textit{``I wish it would generate more than two, like four or more.''}
As text-to-music models often fill in unspecified details, participants added constraints to their prompts to steer results (\textit{e.g., ``no strings,'' ``remove brass''}). 
Many participants found it challenging to rewrite full prompts, wanting instead to provide feedback as \textit{critique}: \textit{``This is too sad and slow. How should I change to make it less sad?''} (P2).
To narrow down candidates, participants marked tracks they liked and unliked. P6 noted that pointing to examples is easier than articulating how to change: \textit{``Picking what I don't like is easy, but describing why I don't like is hard''}.

Even after selecting a favorite, participants kept generating alternatives. P3 explained, \textit{``I still make backups because once I change the cut [in the video editor] or change the overall vibe, I might need a different option.''} Participants tried to steer generation to match the video scene by prompting exact timing or tempo, but found the results did not adhere to the prompt. After edits, they wanted clear before/after comparisons: \textit{``Explicitly state what changed; right now I have to manually check''} (P3).

\ipstart{Organization gets harder as generations pile up}
Participants generated an average of 15 (SD = 5.44) tracks each during the study. As sets grew, managing versions, favorites, and use-cases became non-trivial. To avoid missing good music candidates, P6 listened to all generated tracks from the beginning. P5 found her favorites were spread far apart in the list so she kept scrolling back and forth to compare and decide.

As the default instrumental tracks appeared as ~\textit{``Untitled''}, 3 participants adopted ad-hoc titling for management, but found it inefficient (P2-P3, P7). 
P7 titled by intended placement in the video (\textit{e.g.,} ``podcast intro'', ``outro'') and P3 used brief style-based labels (\textit{e.g.,} ``piano blues'', ``sad harp'') but dropped the practice after two trials, finding it challenging to come up with a title every time.
They also wanted help selecting based on intent and fit, as P3 described: \textit{``I want it [future system] to watch my video and shortlist only the ones that are a good match. So I can prioritize what to listen to.''}

\subsection{Design Goals}

Based on our observations, we distill 5 core design goals for VidTune to harness the creative flexibility of generative music for videos. These goals address key user tasks in music generation: exploration (D1, D2), evaluation (D3), refinement (D4), and organization (D5).

\begin{itemize}
\item[\textbf{D1.}] Provide contextualized music prompt suggestions
\item[\textbf{D2.}] Generate diverse but relevant music options
\item[\textbf{D3.}] Facilitate efficient music evaluation and comparison
\item[\textbf{D4.}] Support iterative refinement through user feedback
\item[\textbf{D5.}] Enable flexible management of generated music alternatives
\end{itemize}




\begin{figure*}[!ht]
  \centering
  \includegraphics[width=\textwidth]{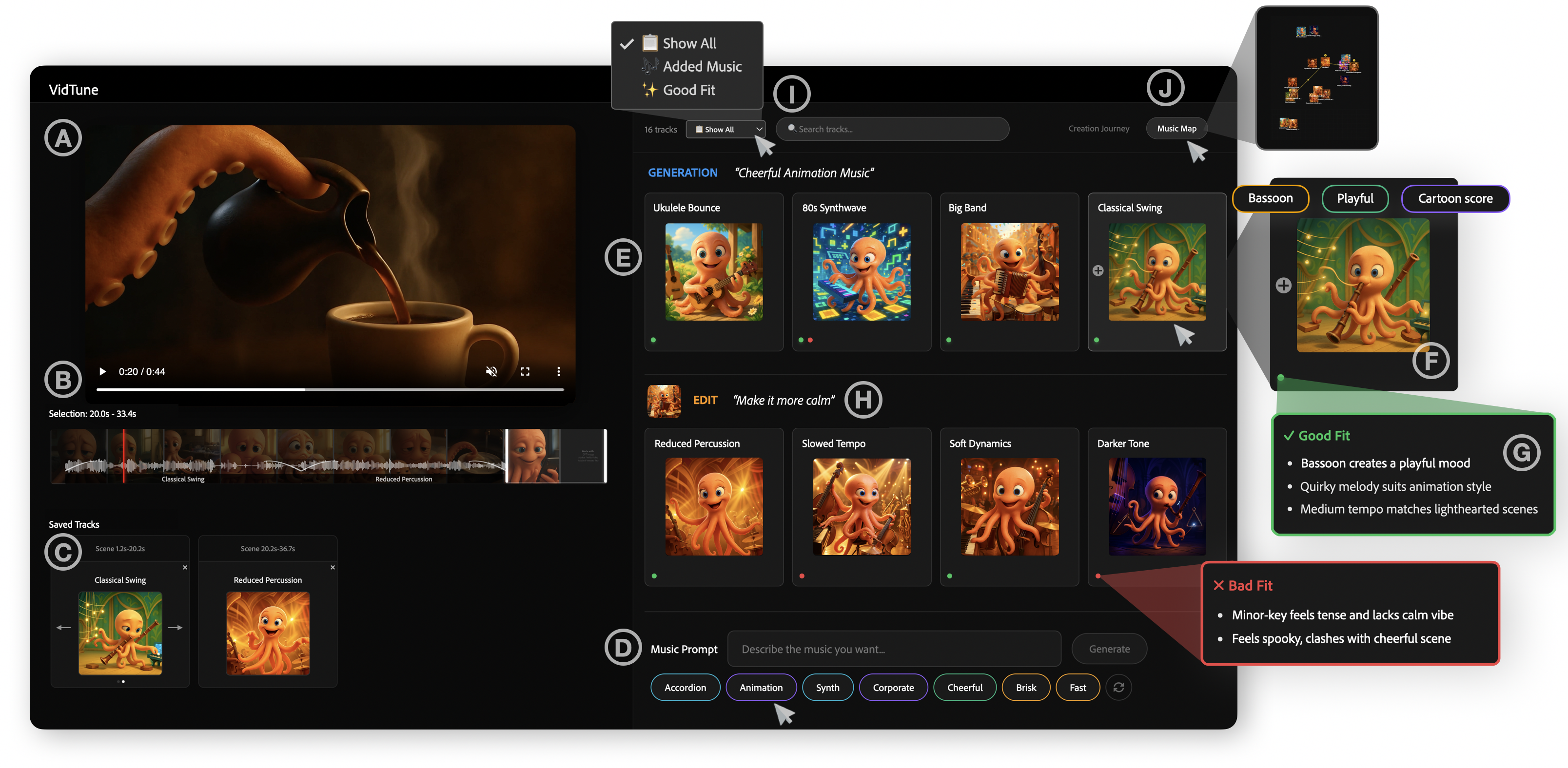}
  \caption{VidTune Interface: The video player and timeline let users choose a scene to add music and preview candidates in sync with the video (A-C). VidTune surfaces prompt suggestions based on the selected scene and user goal (D), then expands the prompt to generate 4 candidates with \textit{contextual thumbnails} (E). On hover, users see reusable prompt keywords (F) and a \textit{fit check} (G). Users can iterate with natural-language-edits (H), organize generations via filter/search (I), or view a \textit{music map} for similarity-based exploration (J).}
  \Description{VidTune dashboard: a dark-themed UI with a coffee-pouring video preview and synced waveform timeline on the left, and on the right a grid of octopus-themed contextual thumbnails for four music candidates with hover tags (e.g., “Bassoon,” “Playful,” “Cartoon score”), a prompt field with chips and Generate button, natural-language edit cards, green “Good Fit” and red “Bad Fit” callouts, filter/search controls, and a mini “Music Map” cluster for similarity exploration.}
  \label{fig:main_interface}
\end{figure*}

\section{VidTune Interface}
\label{sec:vidtune}


Guided by the design goals, we developed VidTune (Figure~\ref{fig:main_interface}), an interactive system that supports video soundtrack generation with text-to-music models. Users can generate diverse music options tailored to their videos (D1, D2), quickly review and compare tracks with contextual thumbnails (D3), and iteratively refine them with \textit{edit}, \textit{vary}, and \textit{blend} interactions (D4). As users generate more tracks, VidTune simplifies management through filtering, searching, and grouping by similarity (D5). Building on Hammad et al.~\cite{hammad2025s}, which shows that creators want to validate music in the context of their video, VidTune integrates a video player. 

\ipstart{Writing prompts with contextual suggestions}
Once the user uploads a video, they enter basic project details: title, video type, target audience, and soundtrack goal. On the left are the video player and filmstrip (Figure~\ref{fig:main_interface}A-B), where users can select scene segments for which to generate music. Based on this selection, VidTune displays \textit{prompt suggestions} (Figure~\ref{fig:main_interface}D) relevant to the current scene and the user-provided project details. Suggested keywords span instruments (\textit{e.g.,} guitar, bass), genres (\textit{e.g.,} jazz, indie pop), and vibes (\textit{e.g.,} upbeat, cheerful). Users can click to add suggestions, refresh to see alternatives, or type custom text. VidTune generates 4 candidate music tracks from a single prompt (Figure~\ref{fig:main_interface}E) by expanding the prompt into four different alternatives and generating one track along with a descriptive title for each. 

\begin{figure*}[!t]
  \centering
  \includegraphics[width=\textwidth]{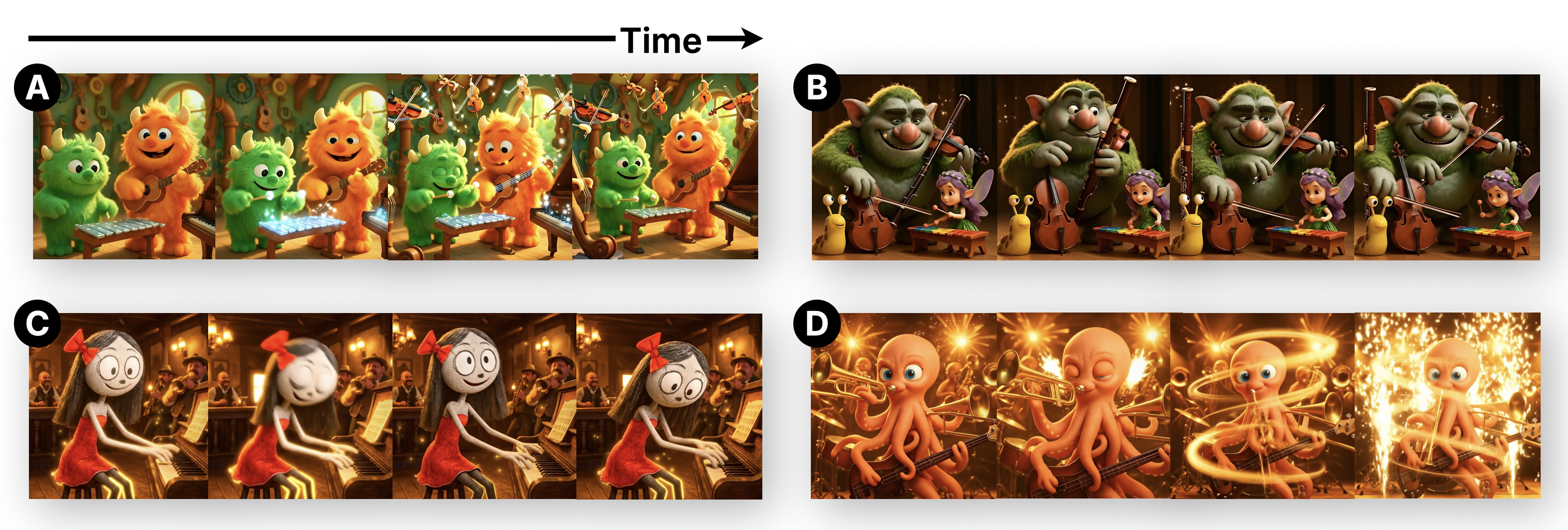}
  \caption{VidTune's animated thumbnail sampled at 0.5 FPS. As the thumbnail animates, the character switches to instruments that enter later in the audio (A, B), movements convey rough tempo (C), and visual effects convey energy (D).}\label{fig:video_thumbnails}
  \Description{ Four time-ordered strips (A–D) of 0.5-FPS thumbnails: A) toy-like monsters rotate to new instruments as parts enter; B) troll ensemble introduces later-arriving strings; C) a cartoon pianist’s head/body bounces to signal tempo; D) an octopus musician gains swirling light rings and sparks to convey rising energy; “Time →” marks left-to-right progression.}
\end{figure*}

\ipstart{Reviewing generations with contextual thumbnails}
Each generated track is paired with a \textit{contextual thumbnail} that anchors on core subjects from the user's footage (\textit{e.g.,} the octopus character) and conveys genre, mood, tempo, and instrumentation through artistic style, color, and motion effects.
Contextual thumbnails let users rapidly skim candidate tracks, spot differences between tracks, and help remember tracks~\cite{childers1984conditions}. 
\revised{As the user plays each track, its thumbnail animates into an 8-second loop (Figure~\ref{fig:video_thumbnails}) making tempo and energy more salient and memorable. Rather than animating all thumbnails at once, we only animate the currently playing track to avoid visual clutter. The static thumbnails still convey tempo and energy through motion cues such as blur and speed lines.}


Each thumbnail includes a \textit{fit check}, a subtle green/red indicator signaling alignment between the generated music and the user's prompt and video. On hover, VidTune explains the rationale (Figure~\ref{fig:main_interface}G). 
Hovering over the thumbnail also reveals \textit{reusable keywords} (Figure~\ref{fig:main_interface}F): prominent attributes specific to the current track that are not already specified in the original prompt or title. Users can click a keyword to add it to the prompt for future generations. 
If the user likes a track, they can save it to the scene with the $\oplus$ icon. 
The track appears in \textit{Saved Tracks} (Figure~\ref{fig:main_interface}C) and its waveform is shown on the filmstrip (Figure~\ref{fig:main_interface}B). VidTune automatically adds short fade-in/out at boundaries for smoother transitions, and the video player previews the video and selected track together. Users can attach multiple alternatives to the same scene and switch between them using the carousel arrows. 

\ipstart{Iterating with natural language}
When users hover over a thumbnail, \textit{Edit} and \textit{Vary} buttons appear.
To edit a track, users can type in a natural language instruction (\textit{e.g., ``Make it more calm''}). VidTune responds with 4 new tracks that implement the user's request in different ways (\textit{e.g.,} reduced percussion, slowed tempo, softer dynamics, or darker tone). Each new track has a title summarizing the change (Figure~\ref{fig:main_interface}H). 
\revised{Clicking \textit{Vary} generates 4 new tracks conditioned on the selected track's audio. These variations typically preserve high-level properties such as genre, tempo, and instrumentation while changing local phrasing and texture, giving users quick alternatives to a track they already like.}

\begin{figure*}[htbp!]
  \centering
  \includegraphics[width=0.7\textwidth]{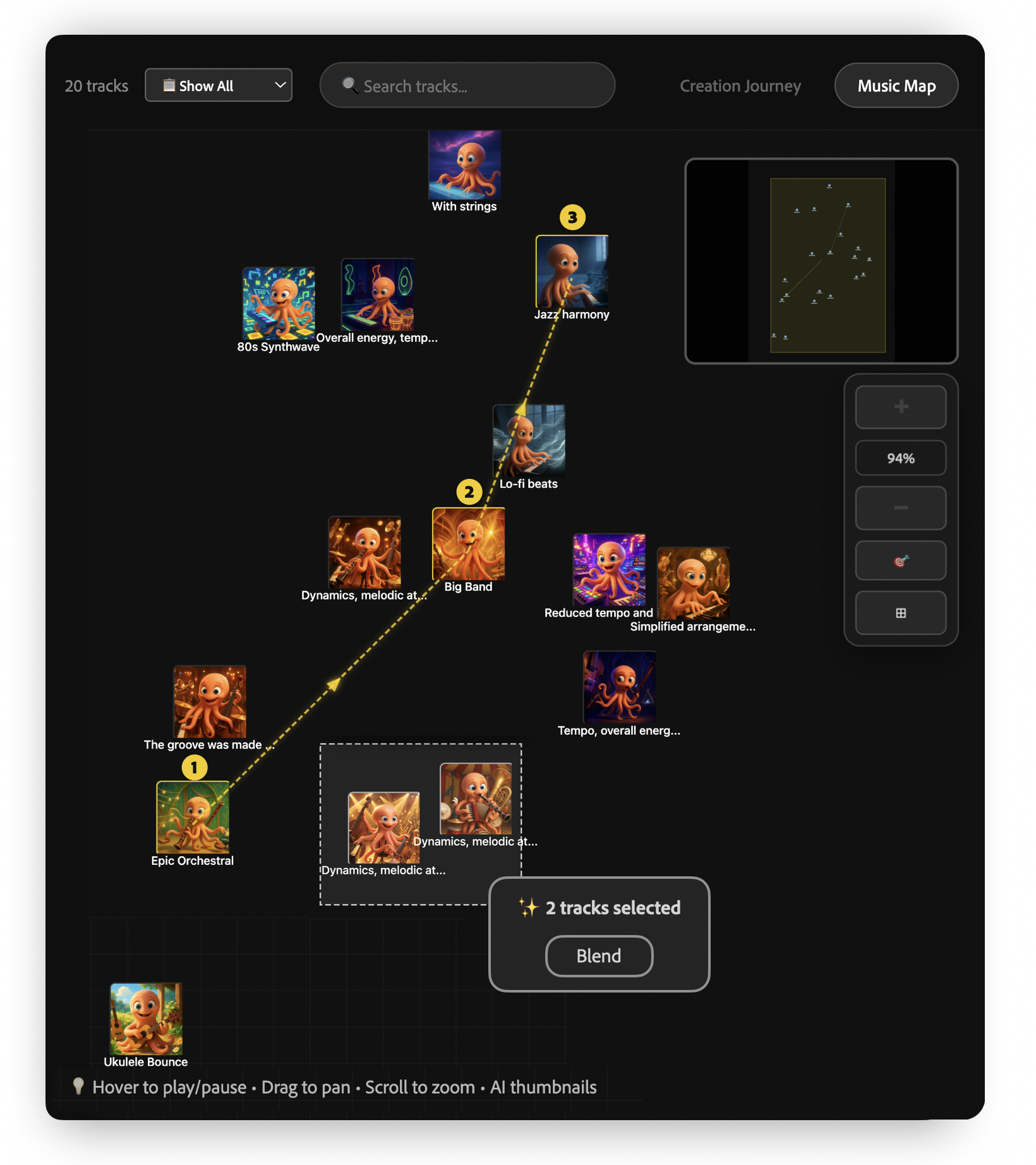}
  \caption{VidTune’s Music Map arranges generated tracks in a 2D space by audio-embedding similarity (CLAP~\cite{wu2023large}), revealing families of related music at a glance. A dashed path shows the sequence of tracks added to the current video. Users can multi-select tracks and use \emph{Blend} to generate similar variations. 
  }\label{fig:music_map}
  \Description{Music Map view: a dark canvas with scattered octopus track thumbnails positioned by similarity, a numbered dashed path (1→2→3) tracing chosen tracks, a dotted multi-select box with “2 tracks selected” and a “Blend” popover, search/filter controls on top, a mini minimap and vertical toolbar (zoom, percent meter, favorite, grid) on the right, and hover-to-preview guidance along the bottom.}
\end{figure*}

\ipstart{Managing generation alternatives}
As generations accumulate, users can scroll through the grid and narrow down options with \textit{filters} (``Show All'', ``Added Music'', ``Good Fit'') or \textit{search} by keywords such as instruments or vibe (Figure~\ref{fig:main_interface}I).
For a broader view, users can switch to the \textit{Music Map} (Figure~\ref{fig:music_map}), which replaces the grid view with a 2D layout that arranges tracks by audio similarity. 
While prior maps for music visualization render nodes as dots~\cite{petridis2022tastepaths} or artist photos~\cite{gajdusek2021spotifygraph}, our contextual thumbnail nodes encode rich audio cues (\textit{e.g.,} instruments, mood), making neighborhoods more interpretable and easier to skim at a glance.
VidTune's map overlays a yellow dashed path indicating the sequence of tracks already placed in the video. 
Users can select nearby tracks to explore close variants or choose multiple tracks and click \emph{Blend} to generate a new option that combines their qualities.

\section{VidTune Algorithms}


\subsection{Suggesting and Diversifying Prompts}VidTune automatically suggests prompts for text-to-music models grounded in the user’s video context, and expands them into diverse, non-conflicting variations.

\ipstart{Suggesting prompts}~\label{sec:pipeline_suggest} 
VidTune first segments the video into short scenes with a Large Multimodal Model (LMM), capturing story changes such as emotions, events, or settings. For each scene, the system generates 5 keywords per category (instruments, genres, vibes, and energy levels, inspired by the categories in~\cite{hammad2025s}), and when users select a region in the filmstrip, a random subset (1–2 per category) is surfaced to support quick prompting. Clicking ``refresh'' shows a different random subset from the pre-analyzed set.
As users begin placing music tracks on the timeline, VidTune uses an LMM to generate a detailed caption of each selected track. These captions are incorporated as additional context for future keyword generation, allowing subsequent suggestions to reflect both the video scene and the user’s emerging preferences. We include the full prompts used for generating the suggestions in the Appendix.

\ipstart{Expanding prompts}
To help users explore diverse possibilities, we designed an algorithm that expands each prompt into \(N\) variations and generates a corresponding set of candidate tracks (Alg.~\ref{alg:prompt-expand}). First, we use an LMM to generate \(N\) diverse, non-conflicting suffix modifiers (2–3 words each) spanning genre, instrumentation, mood, and energy. Each modifier is appended to the user’s original prompt to form a complete prompt for the text-to-music model. For example, given the prompt \textit{``piano solo''}, the system may suggest modifiers such as \textit{``meditative ambient''} or \textit{``jazz swing''}, but not \textit{``guitar''} or \textit{``strings''}, which would contradict the solo-instrument constraint.  

Once candidate tracks are generated, VidTune evaluates their suitability to guide selection. Each track is embedded in CLAP~\cite{wu2023large}, a multimodal model that maps audio and text into a shared representation space for semantic similarity comparisons. We use CLAP embeddings to score each track by (i) \textit{scene fit}, measuring similarity to a vibe description of the scene (derived from the LMM and embedded as CLAP-text), and (ii) \textit{taste fit}, measuring similarity to the mean embedding of tracks the user has previously added to the timeline. We empirically set \(N=6\) to balance variety and generation speed, and present the top 4 ranked tracks to the user.

\begin{algorithm}[h]
\caption{Prompt Expansion}
\label{alg:prompt-expand}
\begin{algorithmic}[1]
\STATE \textbf{Inputs:}
\STATE \hspace*{1em} $Q_t$: user query
\STATE \hspace*{1em} $\mathbf{c}$: scene vibe embedding (CLAP-text)
\STATE \hspace*{1em} $\mathbf{u}$: taste prior (mean CLAP-audio of added tracks)

\STATE \textbf{Generate modifiers:} $\mathcal{S} \leftarrow \mathrm{Gemini}(Q_t)$
{\color{gray}\footnotesize \(N\) non-conflicting suffixes}

\STATE \textbf{Form prompts:} $\mathcal{P} \leftarrow \{\, Q_t \oplus s \mid s \in \mathcal{S} \,\}$

\FOR{$p \in \mathcal{P}$}
  \STATE $a(p) \leftarrow \mathrm{T2M}(p)$ \; {\color{gray}\footnotesize music generation}
  \STATE $\mathbf{z}(p) \leftarrow \mathrm{CLAP\_Audio}(a(p))$
\ENDFOR

\STATE \textbf{Scene fit:} $R(p,\text{scene}) = \cos(\mathbf{z}(p), \mathbf{c})$
\STATE \textbf{Taste fit:} $R(p,\text{taste}) = \cos(\mathbf{z}(p), \mathbf{u})$
\STATE \textbf{Score:} $R(p) = \alpha R(p,\text{scene}) + (1-\alpha)R(p,\text{taste})$
\STATE \textbf{Filter:} $\mathcal{P}' \leftarrow \text{TopK}(\mathcal{P}, 4 \;\text{by}\; R(p))$
\STATE \textbf{Output:} 4 tracks $p \in \mathcal{P}'$, ordered by $R(p)$
\end{algorithmic}
\end{algorithm}

\subsection{Generating Contextual Thumbnails}\label{sec:pipeline_thumbnail}
\begin{figure*}[htbp!]
  \centering
  \includegraphics[width=\textwidth]{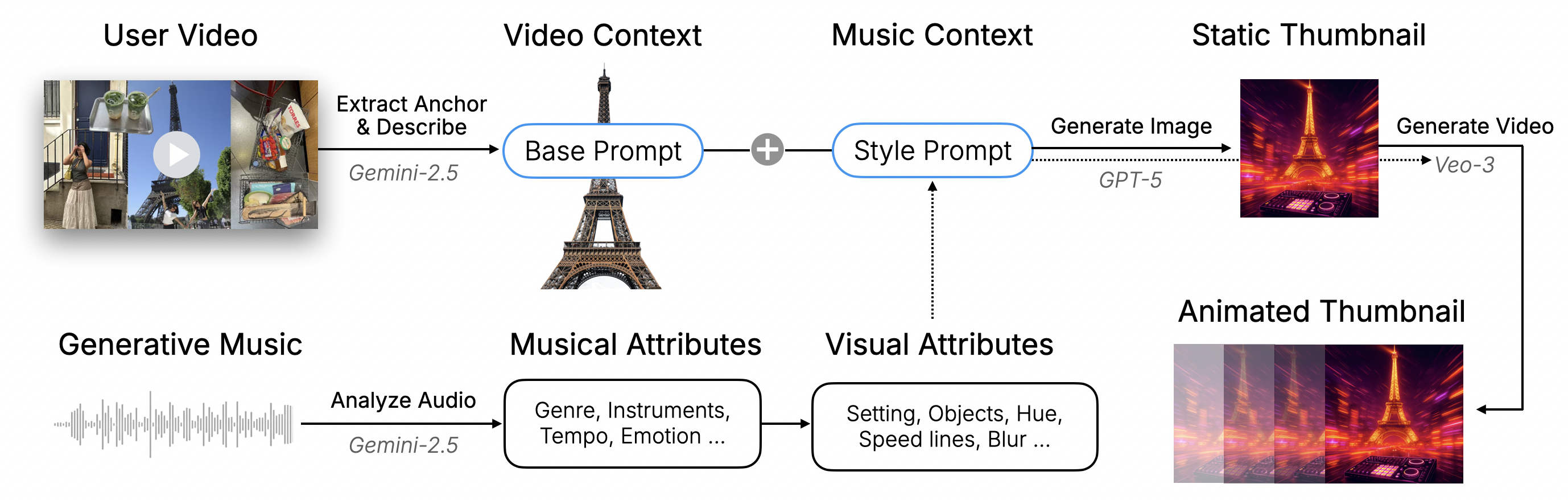}
  \caption{From a user video, VidTune extracts an anchor subject and description to form a \emph{base prompt} grounded in the footage. It analyzes the generated music to infer musical attributes, maps them to visual cues, and fuses these into a \emph{style prompt}. The fused prompt is used to generate static and animated thumbnails that reflect both the video context and the music.}
  \Description{ Pipeline diagram: user video (Eiffel Tower clip and scenes) → Gemini-2.5 extracts an anchor/description to form a Base Prompt; the generated music waveform is analyzed (genre, instruments, tempo, emotion) and mapped to visual cues (setting, objects, hue, speed lines, blur) to form a Style Prompt; Base + Style are fused → GPT-5 creates a static thumbnail → Veo-3 renders an animated thumbnail, with labeled arrows showing data flow between modules.
}
  \label{fig:thumbnail_pipeline}
\end{figure*}

VidTune's contextual thumbnails allow creators to instantly see and compare how different musical options would feel in the context of their actual footage. To generate the thumbnails, VidTune 1) extracts the core semantic visual anchor from the video  (\textit{e.g.,} key subject or scene) and 2) analyzes each track for musical attributes including genre, valence, energy, instruments, and tempo. 
Then, we generate each thumbnail by blending the musical attributes directly into the style of the visual anchor (Figure~\ref{fig:thumbnail_pipeline}).


\ipstart{Identifying the visual anchor}
Our pipeline first uses an LMM to identify at most three core visual anchors in the video. When the anchor is an animated or non-human character (\textit{e.g.,} a cartoon figure, animals), VidTune recreates it as closely as possible to maintain consistency with the video. When the anchor is a real human, instead of reproducing their likeness, VidTune generates a stylized avatar that reflects key appearance details (\textit{e.g.,} hair color, clothing) while avoiding direct modification of real people~\cite{chesney2019deep, lima2021survey} and the associated uncanny valley concerns~\cite{mori2012uncanny}. To maintain consistency across multiple thumbnails, VidTune uses an LMM to generate a post-description of each detected anchor, which is reused to render visually coherent images.
When no clear protagonist is present, the LMM identifies the central object or theme of the scene. For example, in a travel vlog, it detects ``Paris'' as the theme and selects a representative frame (\textit{e.g.,} Eiffel Tower) as the anchor (Figure~\ref{fig:thumbnail_pipeline}). In videos centered on a core object -- such as an advertisement or a product demo -- the object itself is chosen as the anchor.

\begin{table}[htbp]
\centering
\sffamily
\renewcommand{\arraystretch}{1.15}    
\setlength{\tabcolsep}{0.6em}         

\begin{tabular}{p{0.32\columnwidth} p{0.58\columnwidth}}
\toprule
\textbf{Musical Attribute} & \textbf{Visual Mapping Rule} \\
\midrule
Genre \& Style & Background scene and artistic style \newline \textit{e.g.,} electronic → neon cityscape \\
\midrule
Instruments & Protagonist performing instrument (size proportional to prominence) \\
\midrule
Tempo & Implied motion \newline \textit{e.g.,} fast → speed lines \& blur \\
\midrule
Emotion \& Mood & Facial expressions and body language matching emotion \\
\midrule
Valence & Visual filter: hue/tint adjustment \newline \textit{e.g.,} positive → warm, negative → cool \\
\midrule
Energy & Visual filter: brightness and saturation \newline \textit{e.g.,} high → bright, low → dim \\
\bottomrule
\end{tabular}

\vspace{6pt}
\caption{Mapping of musical attributes to AI-generated thumbnails in VidTune.}
\label{tab:music_to_thumbnails}
\Description{Two-column table mapping music traits to thumbnail rules: genre/style → background scene \& art style (e.g., electronic→neon cityscape); instruments → protagonist shown playing, size ∝ prominence; tempo → implied motion (speed lines/blur for fast); emotion/mood → matching facial expressions/body language; valence → hue/tint (warm=positive, cool=negative); energy → brightness/saturation (bright=high, dim=low).
}
\end{table}

\ipstart{Analyzing music and generating thumbnails}
In order to produce visual information from music tracks in a consistent way, we developed a mapping framework (Table~\ref{tab:music_to_thumbnails}) grounded in color–emotion correspondence \cite{valdez1994effects, palmer2013music}, links between tempo/energy and perceived motion speed \cite{davis2018visual}, and affective associations of genre as settings \cite{lehtiniemi2012animated}. Beyond aggregating prior works, this framework operationalizes them into a single, composable schema that yields consistent, multi-attribute thumbnails for music. Whenever a new track is generated, we prompt an LMM to describe its musical character using this schema, conditioning the prompt with the mapping rules and a few examples \cite{brown2020language}.
From this description, we then generate a \textit{stylistic modifier} -- a short phrase that translates the music's qualities into visual effects.  Finally, the stylistic modifier is appended to the visual anchor's description to construct a complete prompt for the text-to-image model. For example, in Figure~\ref{fig:thumbnail_pipeline}, the anchor ~\prompt{`a shot of the Eiffel Tower'} is combined with a modifier derived from upbeat electronic music to become: ~\prompt{`A cinematic shot of the Eiffel Tower at night. The artistic style is futuristic electronic, with the surrounding scene transformed into a vibrant neon cityscape} (Genre \& Style). \prompt{The image is bright and highly saturated with a warm color palette of gold and magenta} (Energy \& Valence). \prompt{There is a dynamic sense of implied motion, with subtle light streaks and motion blur to reflect the fast tempo} (Tempo)'. 
\revised{Finally, we generate an animated thumbnail with a video model that takes the static thumbnail as the first frame and uses the same music-informed style prompt, so that tempo descriptors (\textit{e.g.,} rapid light streaks, slow drifting movement) are reflected in the animation speed and motion intensity.}




\subsection{Representing the Music Space}\label{sec:pipeline_music_map}
\revised{To provide a global view of all generated tracks, VidTune projects them into a two-dimensional music map. Each track is embedded in CLAP~\cite{wu2023large},
and we compute audio–audio similarities between tracks and apply t-SNE~\cite{maaten2008visualizing} to position them so that nearby points correspond to CLAP-similar tracks. In practice, we observed that local neighborhoods tend to group tracks with similar genre, instrumentation, and mood, so we treat the map as an approximate similarity layout for exploration rather than as interpretable axes for specific musical features. As users generate or save new tracks, the map updates dynamically, allowing them to understand explored regions and identify gaps for further exploration.
}

\subsection{Refining Music Generations}\label{sec:pipeline_iterate}
VidTune supports 3 operations for iterative refinement: \textbf{edit}, \textbf{vary}, and \textbf{blend}. 
\textbf{Edit} requests (\textit{e.g., ``make it more energetic''}) are interpreted by the LMM into 4 alternative strategies (\textit{e.g., ``increase tempo,'' ``add percussion,'' ``brighter chord progression''}).
For each strategy, the LMM rewrites the track’s structured music description (from \S\ref{sec:pipeline_thumbnail}) to preserve core attributes while reflecting the requested modification, and uses the rewritten descriptions to regenerate the 4 edited candidates.
\textbf{Vary} regenerates new options by providing the audio of a selected track as input to condition the model, producing close variations to the original.
\textbf{Blend} combines multiple tracks by prompting the LMM with their descriptions to produce a unified prompt capturing their shared traits. This blended prompt is then used to generate 4 new tracks that inherit qualities of both inputs. Together, these interactions help users progressively refine the soundtrack while maintaining coherence across scenes.



\subsection{System Implementation}
We implemented VidTune as a full-stack application. The interface is built in React and TypeScript and uses native Web APIs~\cite{webapis_mdn} for video playback and audio fade-in/out effects. The backend is a Flask REST server~\cite{flask} that processes audio with FFmpeg~\cite{ffmpeg} and runs CLAP~\cite{wu2023large} in PyTorch~\cite{pytorch} for embedding generation and similarity scoring. The algorithms use the LMM \texttt{gemini-2.5-flash}~\cite{team2023gemini} for video analysis (prompt suggestions, thumbnail anchor extraction) and music analysis, \texttt{gpt-5-image}~\cite{openai2024gptimage} for generating thumbnail images, and Veo-3~\cite{google_deepmind_veo3_2025} (\texttt{veo-3.0-fast-\allowbreak generate-\allowbreak preview)} for generating animated thumbnails.
We used a custom text-to-music model trained on licensed instrumental music, which takes either a text prompt or an audio file as input along with desired duration, and generates music with the given duration (VidTune uses the duration of the selected scene).
Model implementations are not our core contribution, and any component could be swapped for an alternative (\textit{e.g.,} a different text-to-music model).

\section{Technical Evaluation}\label{sec:technical_eval}
We conducted technical evaluations of VidTune’s core algorithmic components, focusing on (1) whether prompt expansion increases music diversity and (2) how well contextual thumbnails convey the corresponding music.

\subsection{Diversity of Music from Expanded Prompts}\label{sec:prompt_diversity}
\ipstart{Method}
We built a test corpus of 20 diverse music prompts that generated \textbf{160} music tracks. 
To derive prompts that cover diverse soundtrack contexts, we sampled 5 videos (V1-V5 in Table~\ref{tab:video_materials}). For each video, we selected 2 scenes and authored both a short prompt (\textit{e.g.,} \prompt{`lively busy music'}) and a long prompt (\textit{e.g.,} \prompt{`Upbeat and quirky indie pop, 130 bpm. A driving acoustic guitar and bouncy bassline create a positive, energetic rhythm. Features a bright pizzicato string melody suggesting multitasking.'}) for each scene. 
For each prompt, we generated 8 tracks using the same text-to-music model as VidTune: 4 from the original prompt (\emph{Baseline}) and 4 from LLM-expanded variations (\emph{VidTune}). We embedded audio with CLAP (512-d)~\cite{wu2023large} and computed 2 measures per prompt:
\begin{itemize}
    \item \emph{Mean pairwise cosine distance} within the set (higher {=} broader spread).
    \item \emph{Cluster separation} (higher {=} clearer split into clusters). 
\end{itemize}
For cluster separation, we use $k{=}2$ as a minimal check of whether outputs separate into more than one group, serving as a lightweight, interpretable complement to distance.

\begin{table}[b]
\sffamily\def\arraystretch{1.13}\setlength{\tabcolsep}{0.5em}
\centering
\begin{tabular}{@{}lcccccc@{}}
\toprule
\multicolumn{1}{c}{} & \multicolumn{2}{c}{\textbf{Total}} & \multicolumn{2}{c}{\textbf{Short prompts}} & \multicolumn{2}{c}{\textbf{Long prompts}} \\
 & $\mu$ & $\sigma$ & $\mu$ & $\sigma$ & $\mu$ & $\sigma$ \\
\midrule
\multicolumn{7}{c}{\emph{Pairwise cosine distance (0-2)}} \\
\midrule
Baseline & 0.19 & 0.07 & 0.22 & 0.09 & \textbf{0.17} & 0.05 \\
VidTune  & \textbf{0.22} & 0.10 & \textbf{0.29} & 0.12 & 0.16 & 0.04 \\
\midrule
\multicolumn{7}{c}{\emph{Cluster separation}} \\
\midrule
Baseline & 18.87 & 4.89 & 20.27 & 4.89 & \textbf{17.47} & 3.59 \\
VidTune  & \textbf{20.15} & 5.32 & \textbf{23.77}$^{*}$ & 6.21 & 16.53 & 2.33 \\
\bottomrule
\end{tabular}
\vspace{10pt}
\caption{Diversity metrics by prompt length. \textbf{Bold} indicates the larger mean ($\mu$) within each Baseline–VidTune pair and $^{*}$ indicates $p{<}.05$ with paired $t$-test.}
\label{tab:diversity_results}
\Description{Table comparing Baseline vs. VidTune diversity: pairwise cosine distance (0–2) shows VidTune higher overall (μ 0.22 vs 0.19) and for short prompts (μ 0.29 vs 0.22), with Baseline slightly higher for long (μ 0.17 vs 0.16); cluster separation shows VidTune higher overall (μ 20.15 vs 18.87) and for short (μ 23.77* vs 20.27, *p < .05), while Baseline is higher for long (μ 17.47 vs 16.53); σ values reported in each column.}
\end{table}

\ipstart{Results}
\revised{Overall, we did not find a statistically significant difference in diversity between VidTune and the baseline (Table~\ref{tab:diversity_results}). For \textit{short} prompts, VidTune's diversity was noticeably higher, with significance in cluster separation (cosine: $t$=1.96, $p$=0.079; cluster sep.: $t$=2.42, $p$<0.05). Results for \textit{long} prompts showed no reliable differences between conditions.}
The lack of diversity gains for \textit{long} prompts likely reflects that these prompts already encode rich, specific constraints (instrumentation, tempo, style, mood), and even with well-crafted variations, differences can be buried by the dense specification. 
Yet, as seen in our formative study, users tended to use short abstract prompts to explore breadth and long detailed prompts for specific targets. It is therefore reasonable that VidTune’s expansions most effectively increase variety when user intent is open-ended.



\subsection{Music--Thumbnail Correspondence}\label{sec:pipeline_thumbnail_eval}
To assess how well VidTune’s thumbnails both represent their underlying music and support comparison, we conducted a controlled evaluation guided by the following research questions:
\begin{itemize}
    \item \textbf{RQ1}: How well do VidTune’s thumbnails \textbf{represent} their corresponding music?
    \item \textbf{RQ2}: How well do VidTune’s thumbnails help users \textbf{associate} the right music in the presence of similar alternatives?
\end{itemize}

\ipstart{Baseline}
Generating visual thumbnails to represent music is not new, and commercial tools already use them.
We include this practice as a baseline for comparison, but these tools do not disclose how thumbnails are produced. Our analysis of tracks and their thumbnails generated in the formative study shows that they often loosely reflect explicit prompt terms (\textit{e.g.,} ``rainbow'' leading to rainbow imagery) or the general vibe of the music (\textit{e.g.,} relaxed tracks paired with sky or beach scenes). 
We approximate this by providing both the music prompt and the generated track to an LMM~\cite{team2023gemini} along with instructions to generate an image prompt, which is then used to create a thumbnail image with the same text-to-image model~\cite{openai2024gptimage} as VidTune.

\begin{table}[t]
\sffamily\def\arraystretch{1.13}\setlength{\tabcolsep}{0.5em}
\centering
\begin{tabular}{@{}lcccccc@{}}
\toprule
\multicolumn{1}{c}{} & \multicolumn{2}{c}{\textbf{Total}} & \multicolumn{2}{c}{\textbf{Short prompts}} & \multicolumn{2}{c}{\textbf{Long prompts}} \\
 & $\mu$ & $\sigma$ & $\mu$ & $\sigma$ & $\mu$ & $\sigma$ \\
\midrule
\multicolumn{7}{c}{\emph{Rating (1--7)}} \\
\midrule
Baseline & 4.63 & 0.62 & 4.72 & 1.80 & 4.55 & 1.92 \\
VidTune  & \textbf{4.99}$^{*}$ & 0.74 & \textbf{5.14}$^{**}$
& 1.52 & \textbf{4.83} & 1.56 \\
\midrule
\multicolumn{7}{c}{\emph{Selection (\%)}} \\
\midrule
Baseline & 69.8 & 11.6 & 76.1 & 42.7 & 63.3 & 48.3 \\
VidTune  & \textbf{88.2}$^{***}$ & 7.3  & \textbf{93.5}$^{***}$ & 24.7 & \textbf{83.0}$^{***}$ & 37.7 \\
\bottomrule
\end{tabular}
\vspace{10pt}
\caption{Results for Rating and Selection by prompt length (within-subject, $n{=}20$). 
\textbf{Bold} indicates the larger mean ($\mu$) within each Baseline–VidTune pair. 
Significance is denoted as $^{*}p < 0.05$, $^{**}p < 0.01$, and $^{***}p < 0.001$ (paired $t$-test).}
\label{tab:thumb_results_len}
\Description{Table of user ratings (1–7) and selection rates (percent) by prompt length showing VidTune > Baseline overall and for short/long prompts—ratings: 4.99±0.74 vs 4.63±0.62 (short 5.14±1.52 vs 4.72±1.80; long 4.83±1.56 vs 4.55±1.92); selection: 88.2±7.3 vs 69.8±11.6 (short 93.5±24.7 vs 76.1±42.7; long 83.0±37.7 vs 63.3±48.3), with *, **, *** marking p<.05, p<.01, p<.001.
}
\end{table}


\ipstart{Method}
We collected \textbf{1{,}600} judgments from 20 annotators (80 trials each: 40 ratings, 40 selections). Stimuli comprised 40 audio clips (20 prompts × 2 variations) and 80 thumbnails (VidTune + baseline per clip). 
We reused the same 5 videos and 20 prompts (short and long) from \S\ref{sec:prompt_diversity} and the videos covered diverse types of anchors for thumbnail generation (\textit{e.g.,} animated characters, landmarks, human avatars). For each prompt, we randomly sampled 2 of the 4 VidTune-generated clips as the audio stimuli. For each clip, we produced a VidTune thumbnail and a baseline thumbnail.\footnote{Study materials and annotation interface details provided as supplementary materials.} Including both short and long prompts enabled the selection task to probe associations under easier (more distinct) and harder (more similar) audio conditions.

Annotators self-identified as experts (N=1), intermediate (N=11), and novices (N=8) in music. We used a within-subjects design with counterbalancing, and the task order was randomized. Thumbnails were shown without video context to test whether our thumbnail alone establishes strong music–image correspondence. Annotators completed:

\begin{itemize}
    \item \textbf{Rating (RQ1):} Given one audio clip and one thumbnail (VidTune or baseline), rate on a 1–7 scale how well the image represents the audio (7 = most representative).
    \item \textbf{Selection (RQ2):} Given one audio clip and two thumbnails, choose the image that matches the audio (2AFC). Both thumbnails were from the same condition (either VidTune or baseline) with one from the target audio and one foil from a different audio variation of the same prompt, mirroring typical use.
\end{itemize}

\setlength{\fboxrule}{0pt}
\begin{figure*}[!t]
  \centering
  \framebox{\includegraphics[alt={},width=0.95\textwidth]{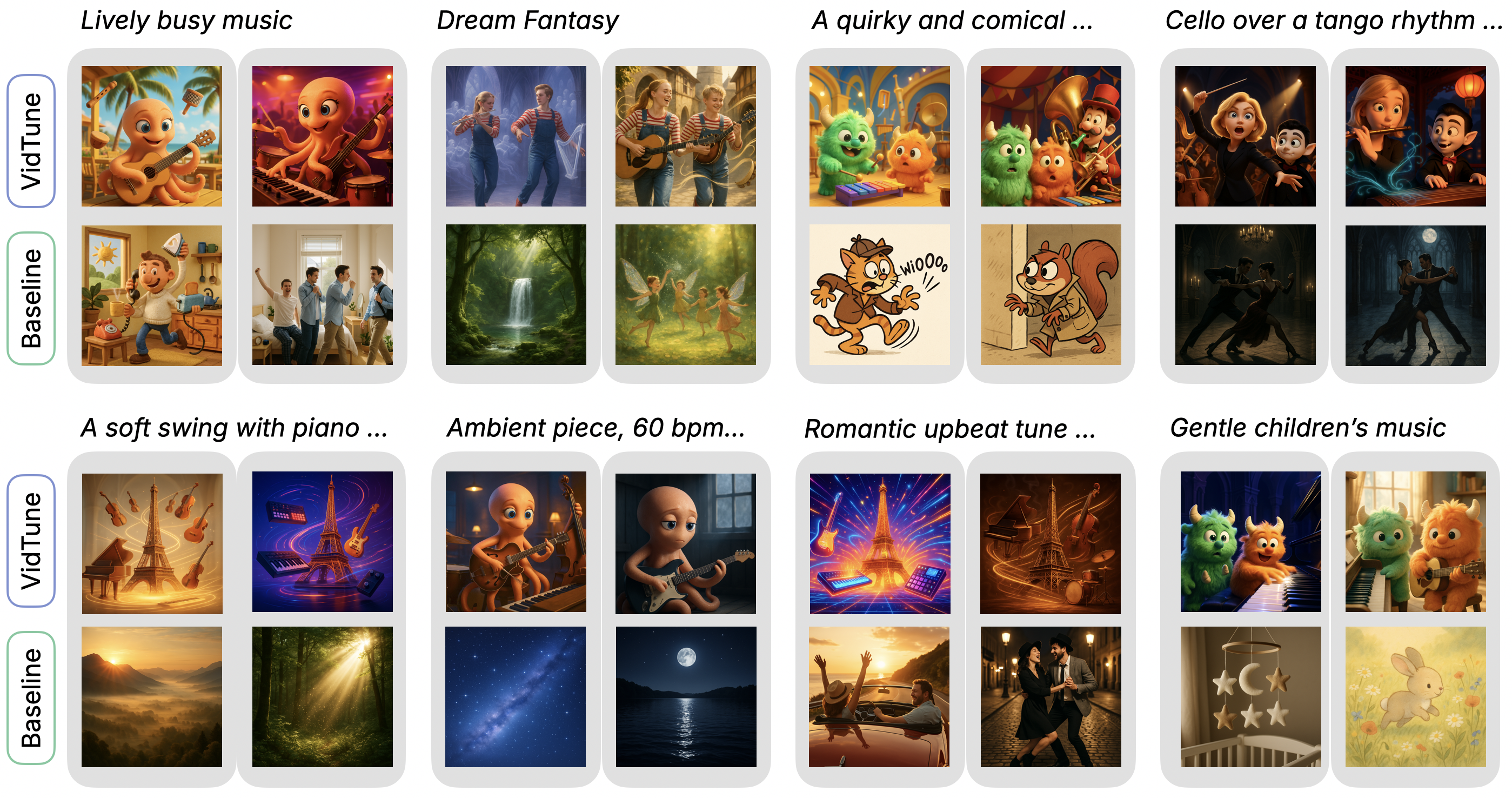}}
  \caption{Example thumbnails from the pipeline evaluation study. For each prompt (column), the top row shows VidTune thumbnails and the bottom row shows the baseline. VidTune tends to convey music‐relevant cues (\textit{e.g.,} genre, instruments), while the baseline leans toward literal prompt imagery or abstract scenes. }
  \label{fig:pipeline_qual_examples}
  \Description{ Eight-column grid comparing VidTune (top) vs Baseline (bottom) thumbnails across prompts—lively busy, dream fantasy, quirky/comical, cello-tango, soft swing with piano, ambient 60 bpm, romantic upbeat, gentle children’s—where VidTune depicts performers/instruments and tempo/energy cues, while Baseline favors literal scenes or abstract/stock-like imagery.}
\end{figure*}

\ipstart{Results}
Table~\ref{tab:thumb_results_len} shows that VidTune thumbnails outperformed the baseline on both the \emph{Rating} and \emph{Selection} tasks. We analyzed within-subject differences using paired \emph{t}-tests and confirmed all effects with Wilcoxon signed-rank tests. 
In the rating task, VidTune thumbnails scored significantly higher than the baseline ($t(19)=2.28$, $p=0.035$, $d=0.51$), with 13 of 20 annotators preferring them. Differences in preference may stem from annotators judging thumbnails without the original video context or preferring abstract illustration over concrete cues (\textit{e.g.,} explicit instruments).
In the selection task, VidTune thumbnails yielded significantly higher 2AFC accuracy than the baseline, with a large effect size ($t(19)$=7.14; $p{<}0.001$; $d$=1.60).
Every annotator performed as well or better with VidTune thumbnails, demonstrating that VidTune's thumbnails reliably encode music and surface distinctions, even when the two audio variants were generated from the same prompt.

Figure~\ref{fig:pipeline_qual_examples} illustrates qualitative examples from VidTune and the baseline as presented to annotators. For VidTune’s thumbnails, differences between candidates were often more distinguishable, conveyed through color palettes, background settings, facial expressions, and depicted instruments.
In the baseline thumbnails, prompts with salient visual components (\textit{e.g.,} tango rhythm, comical) were often literalized, overshadowing musical attributes. Of the 40 baseline thumbnails, only 1 featured instruments present in the audio, 28 featured a character or central object, and 12 were abstract scenes. Distinctions were slightly clearer for \textit{short} prompts (which produced more divergent audio). For instance, the prompt \textit{``Romantic upbeat tune''} yielded one electronic-leaning and one jazzier variation, and the two baseline thumbnails slightly captured that split in mood. 

\section{\revised{Controlled User Study}}\label{sec:user_eval}
To further understand how VidTune supports soundtrack generation, we conducted a within-subjects study with 12 video creators comparing VidTune to a baseline. 



    
    

\subsection{Method}
\ipstart{Participants}
We recruited 12 participants with diverse video creation experience using mailing lists (P9-P20, \S{A} Table~\ref{tab:participants}). 5 described themselves as proficient (P11, P14-P16, P18), having an average of 14 years of video editing experience (SD=6.46), and 7 were amateurs (P9-P10, P12-P13, P17, P19-P20) with 4.86 years (SD=2.27) of experience. The study lasted 1.5 hours, conducted either remotely via Microsoft Teams/Zoom or in-person based on participant preference, and we compensated \$60.

\begin{figure}[t!]
  \centering
  \includegraphics[width=\columnwidth]{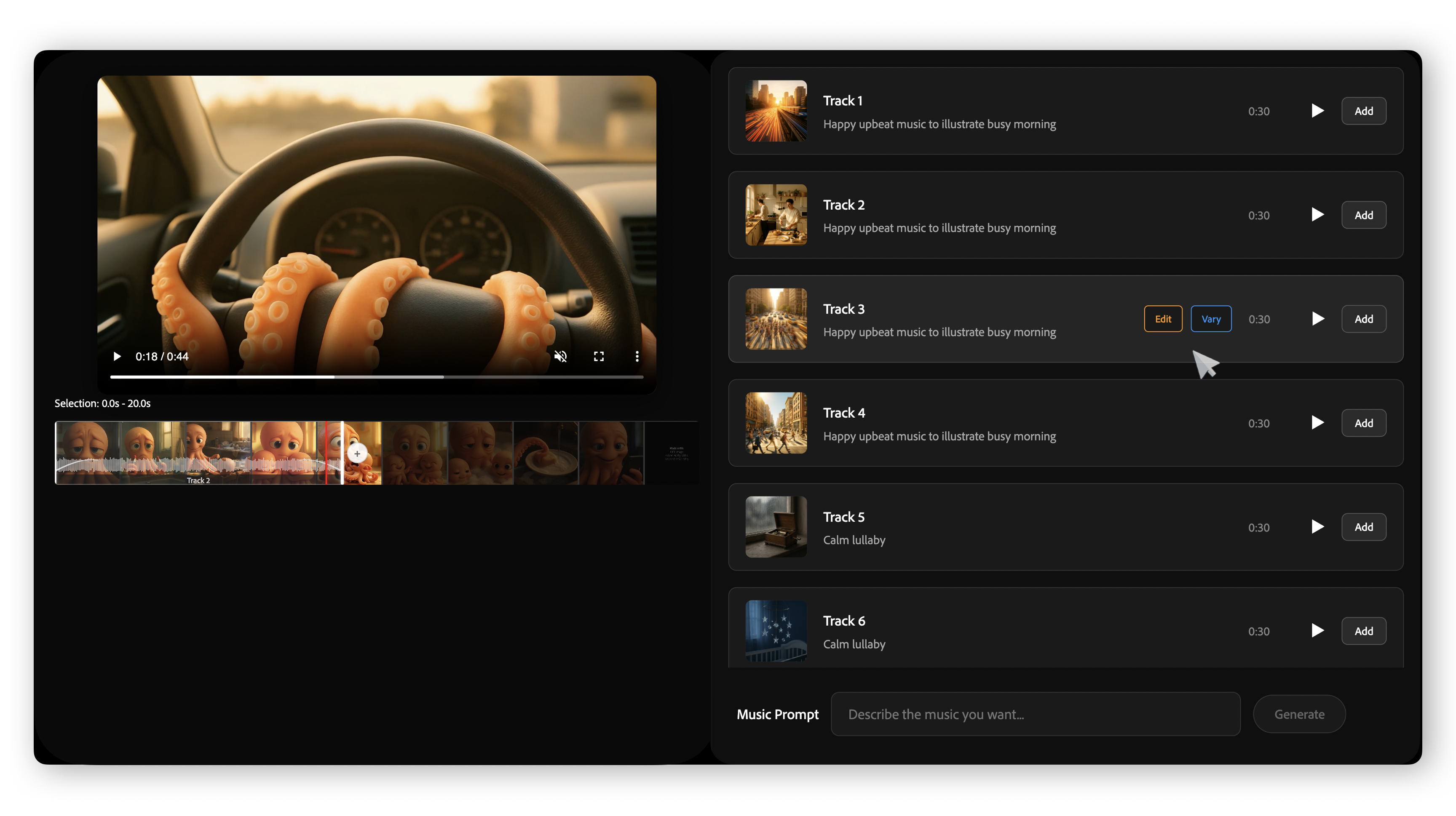}
    \caption{Baseline interface, designed to resemble the UI and features of existing music generation tools.}\label{fig:baseline_interface}
    \Description{ Baseline music-gen UI: left a large video player (octopus tentacles on a car steering wheel) with a filmstrip + waveform selection bar; right a scrollable list of track cards with thumbnails, titles, 0:30 durations, Play/Add and Edit/Vary controls; a single text prompt box with “Generate” sits at the bottom.}
\end{figure}

\ipstart{Baseline}
\revised{To examine the benefits of VidTune relative to existing designs in text-to-music generation, we implemented a baseline (Figure~\ref{fig:baseline_interface}) inspired by widely used commercial systems~\cite{suno, udio}. Because these tools do not document whether they diversify prompts internally, we approximate their typical observable user experience: a single free-form text prompt that yields multiple tracks and simple cover images. For controlled comparison, both the baseline and VidTune use the same underlying music model, generate four tracks per prompt, and share the same video player pane and timeline so users can preview music in sync with their video and see tracks with fade-in/out transitions.}

For thumbnails, we use the baseline generation pipeline from our pipeline evaluation (\S\ref{sec:pipeline_thumbnail_eval}), which conditions an LMM on the user prompt and generated music and then calls the same text-to-image model~\cite{openai2024gptimage} as VidTune. \revised{This produces generic \textit{vibe-oriented} thumbnails without video context or explicit mappings from musical attributes, which are specific to VidTune’s contextual thumbnails. 
The baseline also supports iterative refinement with \textit{edit} and \textit{vary}: \textit{vary} uses the same audio-seed variation mode as VidTune, while \textit{edit} creates a single new prompt with an LLM from the original prompt and edit request and then generates four tracks, but without VidTune’s structured diversification.}

\ipstart{Materials}
We selected 3 videos (V1, V2, V6 in \S{A} Table~\ref{tab:video_materials}) from an independent filmmaker’s short animated films, all generated using tools including Adobe Firefly~\cite{adobe_firefly_2025}, GPT~\cite{openai2024gptimage}, and Veo 3~\cite{google_deepmind_veo3_2025}. Each video was between 45–60 seconds long without narration. V1 was used in the tutorial session for both VidTune and baseline conditions. The main study used V2 and V6, which were similar in length, visual complexity, and scene changes. 

\ipstart{Procedure}
We began with an interview on participants’ video creation practices and typical approaches to adding music. Participants then received a 10-minute tutorial on both VidTune and the baseline interface (V1). Each participant used both interfaces, with up to 25 minutes per interface, to add music to one of two study videos (V2, V6). The assignment of videos and the order of interfaces were counterbalanced and randomly assigned. After each session, participants completed a post-stimulus survey measuring cognitive load (NASA-TLX~\cite{hart1988development}), creativity support (CSI~\cite{cherry2014quantifying}), and custom items related to visual thumbnails. We omitted inapplicable items (\textit{e.g.,} physical demand) and overlapping ones across scales (\textit{e.g.,} NASA-TLX \emph{Effort} vs. CSI \emph{results worth the effort}).




\section{Results}
\begin{figure*}[t]
  \centering
  \includegraphics[width=\textwidth]{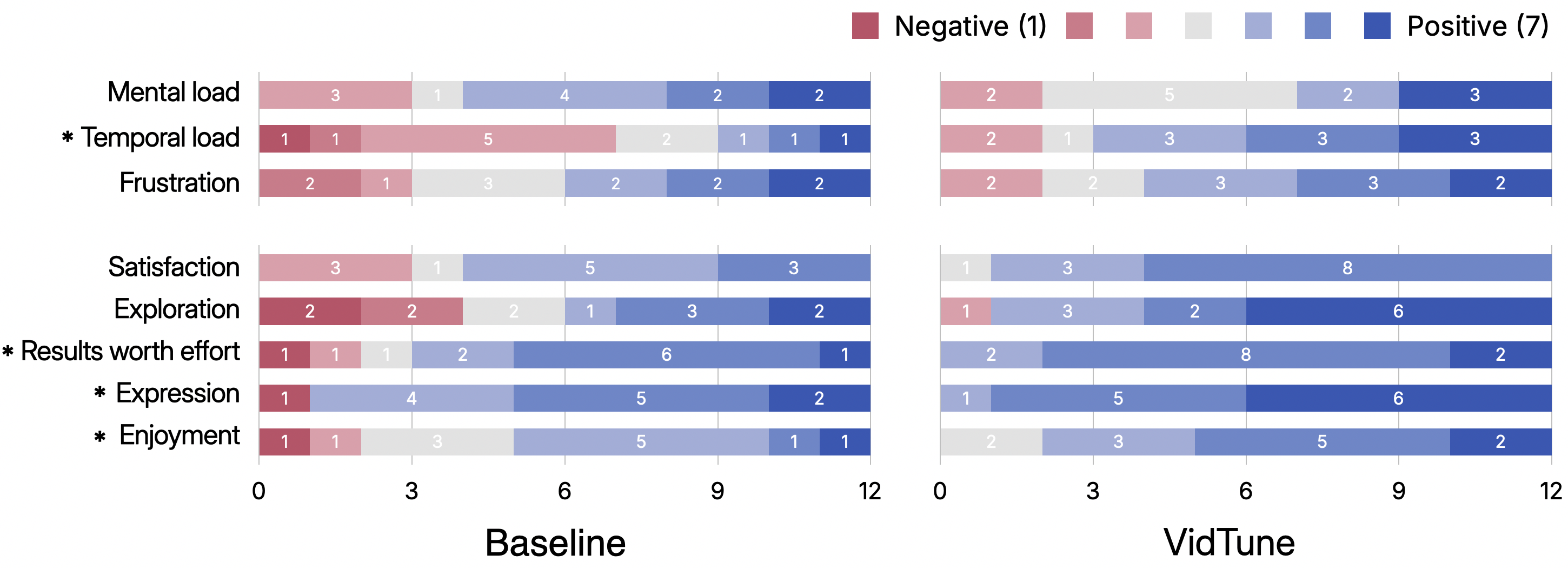}
    \caption{Distribution of rating scores for the Baseline and VidTune (1 = negative, 7 = positive) in the two tasks on cognitive load and creativity support. Higher values indicate more positive feedback. * indicates statistical significance from Wilcoxon tests (\textit{p} < 0.05).}

  \label{fig:survey_results}
  \Description{two side-by-side stacked Likert bar panels (1→7, pink→dark blue) comparing Baseline vs VidTune across mental load, temporal load*, frustration, satisfaction, exploration, results worth effort*, expression*, and enjoyment*; VidTune bars cluster toward higher scores (more dark blue), indicating lower load and higher creativity support, with asterisks marking significant gains.}
\end{figure*}

Figure~\ref{fig:survey_results} summarizes ratings for cognitive load and creativity support, and Figure~\ref{fig:thumbnail_ratings} reports thumbnail-specific ratings. 
In this section, we synthesize findings on VidTune's exploration support (\S\ref{sec:results_exploration}), review support (\S\ref{sec:results_review}), and additional observations on how participants work with VidTune (\S\ref{sec:results_additioanl}).

\subsection{Exploration Support}\label{sec:results_exploration}
On average, participants created 28 tracks in VidTune (SD = 10.65; 1.83 edits, 1.80 variations, 0.17 blends) compared to 16 tracks in the baseline (SD = 1.82; 1.67 edits, 1.00 variations).

\ipstart{Guided exploration reduces effort}
When generating music with VidTune, all participants utilized prompt suggestions and often added their own keywords or explanations to refine the prompt. 
As music generation required waiting time in both conditions, participants often started new generations in parallel or revisited earlier outputs.
Participants using VidTune reported significantly higher result-worth-effort ($\mu$=6.00, $\sigma$=0.60 vs. $\mu$=5.08, $\sigma$=1.68; $Z$=-2.08; $p$<0.05).
They explained that VidTune’s expanded prompts produced more diverse outputs, helping them explore possibilities and find tracks that matched their goals, whereas the baseline often returned 4 very similar tracks, leading them to generate more options before reaching a desirable one. 3 participants also noted that VidTune made iteration easier by allowing them to easily reuse prompts by clicking on keywords (P8, P12-P13).

\ipstart{Requests for expanding expressive range} 
\revised{We did not observe statistically significant differences in perceived exploration support or satisfaction between VidTune and the baseline.}
4 participants wanted more fine-grained control, such as reusing only part of a generated track (P14), extracting and editing stems directly (\textit{e.g.,} only increasing piano volume) (P13, P15, P18), or using more advanced fade types (P18). 
3 participants also input sound effects based on the video scene (\textit{e.g.,} keyboard typing, traffic noises), but noticed the model omitted them (P11-P13). 
2 participants noted that outputs sometimes missed the requested style and wanted brief system explanations on what failed and how to steer it, so they could redirect generation (P17, P19).



\subsection{Contextual Thumbnails for Reviewing Music}\label{sec:results_review}
\begin{figure}[t]
  \centering
  \includegraphics[width=\columnwidth]{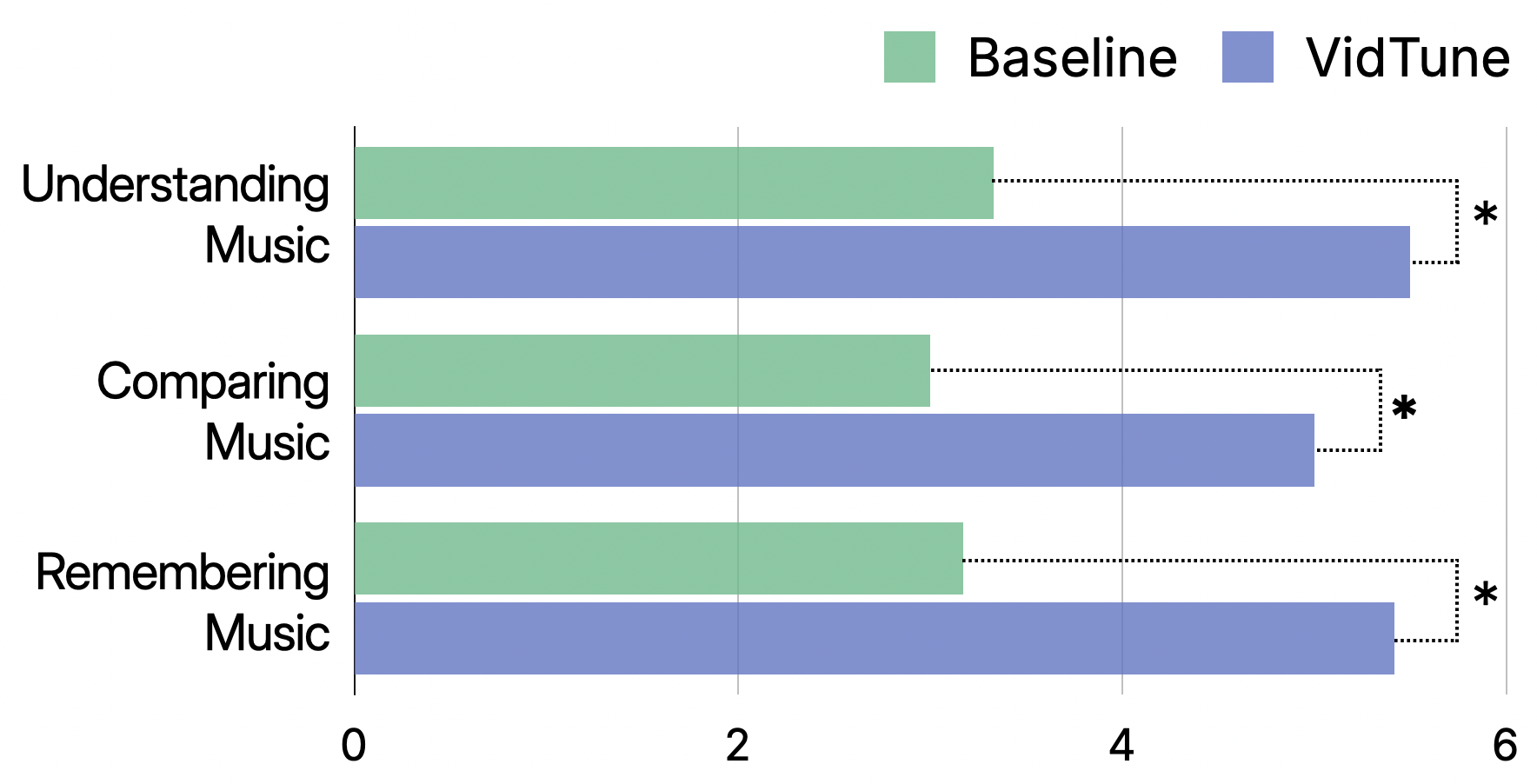}
  \caption{Rating scores on thumbnail-specific questions for the Baseline and VidTune (1 = negative, 7 = positive) across the two tasks. Higher values indicate more positive feedback. Asterisks denote statistical significance from Wilcoxon tests (*~\textit{p} < 0.05).}
  \label{fig:thumbnail_ratings}
  \Description{Horizontal bar chart with three rows—Understanding, Comparing, and Remembering Music—showing Baseline (green) vs VidTune (blue) ratings on a 1–7 scale; VidTune bars extend farther right in all rows, with dashed comparison brackets and asterisks indicating statistically significant improvements.}
\end{figure}

Figure~\ref{fig:thumbnail_ratings} shows that participants found VidTune's thumbnails significantly more useful for understanding music ($\mu$=5.5, $\sigma$=1.51 vs. $\mu$=3.33, $\sigma$=1.78; $Z$=-2.38; $p$<0.05), comparing different tracks ($\mu$=5, $\sigma$=1.41 vs. $\mu$=3, $\sigma$=1.71; $Z$=-2.47; $p$<0.05), and remembering them ($\mu$=5.42, $\sigma$=1.51 vs. $\mu$=3.17, $\sigma$=1.8; $Z$=-2.33; $p$<0.05).

\ipstart{Contextual thumbnails accurately represent music}
Participants described a close match between the music and VidTune's thumbnails, as P16 explained: \textit{``When a thumbnail had a dark background and fireworks, it matched the music really well, so I’d say it was pretty accurate.''} 
P17 noted that thumbnails can sometimes be more helpful than titles in understanding music: ~\textit{``Terms like Celtic jazz or Synthwave are hard to grasp when I just read but much easier when I see them visually.''}
P9 found that the baseline's thumbnails \textit{``conveyed only the `vibe' of a song,''} while VidTune's thumbnails \textit{``also communicate an `acoustic representation' ''}, showing that VidTune further helps anticipate genre, instrumentation, and intensity. 

\ipstart{Contextual thumbnails aid review burden}
VidTune required significantly less temporal demand (Figure~\ref{fig:survey_results}, $\mu$=5.33, $\sigma$=1.44 vs. $\mu$=3.67, $\sigma$=1.67; $Z$=2.13; $p$<0.05). While both systems had similar end-to-end generation times -- and VidTune had additional delay for prompt expansion and music analysis -- participants still experienced VidTune as less rushed. 
With VidTune, 3 participants often skipped tracks and selectively listened to tracks with thumbnails that visually resembled ones they had already liked (P9, P16, P21);
\textit{``because I can tell which would fit my video better by looking at them''} (P9).
In contrast, all 12 participants in the baseline condition listened to all generated tracks, often replaying them because it was hard to visually preview and filter options. P12 explained, \textit{``When it [Baseline] gives many songs that look the same, I feel like I must review all of them.''} showing that VidTune reduced this burden by presenting more distinct thumbnails. Similarly, P10 noted about the baseline thumbnails \textit{``Sometimes it's not what I imagined when I clicked to listen.''}

\ipstart{Contextual thumbnails serve as reliable memory cues}
Participants noted that VidTune's thumbnails support recall~\cite{childers1984conditions}. P11 stated \textit{``It's 100\% helping when finding what music I liked. Names [titles] are not that useful, thumbnails are more useful.''} 
By contrast, baseline thumbnails offered little support for comparison or recall: \textit{``Unless I really pay attention, it’s hard to remember what music I listened to. It's too simple''} (P11).
Additionally, inconsistent visual styles in the baseline further challenged comparing tracks using thumbnails: \textit{``It's a bit too random. Some are illustrations, some are real photos. I cannot compare when one image is showing a group of people partying and the other shows an animated cat''} (P9).
With abstract imagery in baseline thumbnails, participants often forgot which thumbnail matched which track and had to re-listen to identify it.


\ipstart{Contextual thumbnails bring a more engaging, personalized experience}
Participants enjoyed using VidTune significantly more ($\mu$=6.42, $\sigma$=0.67 vs. $\mu$=5.42, $\sigma$=1.56; $Z$=-2.13; $p$<0.05) and felt more expressive ($\mu$=5.58, $\sigma$=1.00 vs. $\mu$=4.50, $\sigma$=1.51; $Z$=-2.13; $p$<0.05).
Although our design goals did not explicitly aim to increase enjoyment, participants described VidTune’s thumbnails as making the music creation process more enjoyable and personally meaningful. 
P17 remarked, \textit{``More pleasing, more fun to interact with. Usually with video editing, I always just look at randomly colored blocks, waveforms, and it’s more like a task. But this makes the process itself more interesting.''} P11 similarly emphasized the sense of personal connection: \textit{``It helps because it uses my characters, feels more connected. Feels more like this is generating music for MY own video.''}

With animated thumbnails, 9 participants found they offered an additional sense of playfulness; as P14 mentioned, ~\textit{``Not sure how professional editors would think, but as a novice creator, this [animation] is just more pleasing.''} Similarly, P16 said they were ~\textit{``more memorable as they're more dynamic.''} 2 participants also noticed additional instruments in the background when the thumbnail animated, and felt that the motion aligned closely with the energy of the track (P10, P16). 
However, 3 participants felt the additional motion did not add much to the static image, describing it as an extra but not essential (P9, P11) or even distracting (P20).

\ipstart{Limitations}
Despite these benefits, participants also pointed out several limitations of VidTune’s thumbnails. 2 participants wanted the option to hide thumbnails and see more detailed textual descriptions of each track, as P20 explained, \textit{``I’m not a visual person, I always prefer to read things.''}
P18 observed that the thumbnails sometimes contradicted the titles, which caused confusion. For example, a track titled \textit{Ukulele Folk} was paired with a thumbnail showing a xylophone. The actual audio had sound qualities that could be interpreted as either instrument, as music generation models often blend overlapping acoustic features~\cite{agostinelli2023musiclmgeneratingmusictext}.
Similarly, P12 mentioned that recurring characters in the thumbnails caused them to look similar. Comparison became harder after multiple rounds of prompt iterations, when the generated samples started to converge and looked nearly identical.

\ipstart{Broader use cases}
Participants suggested additional creative use cases for VidTune's thumbnails.
P15 envisioned using them as a ~\textit{visual scaffold} for iteration ~\textit{``Sometimes I only remember instruments by the visual and not the name, so I want to draw on [VidTune’s thumbnails] or paste [images of] instruments on it.''}
4 participants wanted to reuse them as creative assets -- for example, P18 wanted to export thumbnails for teasers and ending credits of his videos, or use them as emojis with music for fun communication. P9 and P14 mentioned their potential as music videos or album covers for low-budget artists. 
Others highlighted educational and collaborative potential: P9 noted that combining audio with visual thumbnails could help children learn music \textit{``like concept cards''}, and P16 explained that thumbnails could make it easier to communicate musical choices with deaf collaborators. 

\subsection{Additional Observations}~\label{sec:results_additioanl}

\ipstart{Fit checks are adopted selectively}
We observed varied use of VidTune’s fit check feature in participants’ workflows. 3 participants actively relied on fit checks, using them to filter out weaker candidates (P10) or to reassure themselves after adding a track to the video (P9, P12). In contrast, some chose not to use them at all, as P13 explains \textit{``I didn’t read the fit check because it might bias me. Even if the music isn’t bad, if it tells me it’s a bad fit I might not consider it.''} Similarly, P16 noted, \textit{``Fit checks don’t really influence my decision. At the end, it’s more of a creative choice, so there isn’t a right or wrong.''} 

\ipstart{Music map facilitates cross-music consistency}
In both conditions, all but one participant (P20) added multiple tracks to the final video and emphasized the importance of maintaining consistency to avoid abrupt transitions between scenes. 
3 participants used the music map to spot gaps between tracks added to the video (P8, P15, P16). They verified that their tracks were not too distant from one another and compared the thumbnail styles of nearby alternatives in the map. When a candidate felt off, they switched to a closer neighbor or a previously-saved track. P16 used the map's blend feature to generate middle-scene music from the first and last tracks. 


\ipstart{Workflows are iterative}
Participants shaped their soundtracks through small, repeated adjustments. First, they iterated within a scene by requesting changes to mood, speed, and instrumentation (adding or removing) in both conditions. In the baseline, P18 tried repeating the original prompt as an edit when the first generation missed key elements. In VidTune, P12 tried prompting \textit{``add [instrument] from option 2''} but found the system did not support referencing attributes from another candidate.
After choosing one track, participants iterated to generate a next track that kept style consistent while still allowing emotion to shift. 

We observed two iteration strategies: (1) reusing prompt keywords with small, context-specific tweaks (\textit{e.g.,} emotion, tempo, instrumentation), and (2) starting from a favored track, making quick edits or variations, and reusing those across later scenes. Two participants using \textit{Vary} (P14, P15) sometimes struggled to choose among similar options, and wanted clearer text descriptions of differences. Based on these observations, we speculate that the lack of significant differences in mental load (VidTune: $\mu$=4.75, $\sigma$=1.48; baseline: $\mu$=4.92, $\sigma$=1.44; $Z$=-0.31; $p$=0.75) and frustration (VidTune: $\mu$=5.08, $\sigma$=1.38; baseline: $\mu$=4.58, $\sigma$=1.73; $Z$=0.68; $p$=0.50) reflects the need for iterative generation and review, as well as occasional misalignment between steering and intent.

\section{Exploratory Case Study}
\revised{The controlled user study (\S\ref{sec:user_eval}) examined how VidTune compares to a baseline on pre-selected videos.
To understand how VidTune supports soundtrack creation across a more diverse set of personal projects, we further conducted an exploratory case study with creators using their own footage (Table~\ref{tab:exploratory_video_materials}).}


\begin{table}[b]
\sffamily\def\arraystretch{1.0}\setlength{\tabcolsep}{0.5em}
\small
  \centering
  \begin{tabular}{ccccc}
    \toprule
    \textbf{PID} & \textbf{VID} & \textbf{Video Type} & \textbf{Duration} & \textbf{Thumbnail Anchor} \\ \midrule
    P8  & V6 & Vlog (animal)          & 1:01  & Animal \\
    P11 & V7 & Vlog (scenic views)    & 1:05 & Landscapes \\
    P16 & V8 & Vlog (talking head)   & 3:41 & Speaker avatars \\
    P20 & V9 & Product demo            & 1:36  & Mascot \& Product \\
    P21 & V10 & Child Storytelling      & 3:09 & Sketched character \\
    P22 & V11 & Generated Ads           & 1:00  & Characters \& Product \\
    \bottomrule
  \end{tabular}
  \caption{Videos used in the exploratory case study.}
  \label{tab:exploratory_video_materials}
  \Description{Table listing six case-study videos (P8–P22; V6–V11): vlog—animal (1:01), scenic views (1:05), talking heads (3:41); product demo (1:36); child storytelling (3:09); generated ads (1:00); with thumbnail anchors: Animal, Landscapes/Landmarks, Speaker Avatars, Mascot & Product, Sketched Character, Characters & Product.
}
\end{table}

\subsection{Method}
\begin{figure*}[htbp!]
  \centering
  \includegraphics[width=\textwidth]{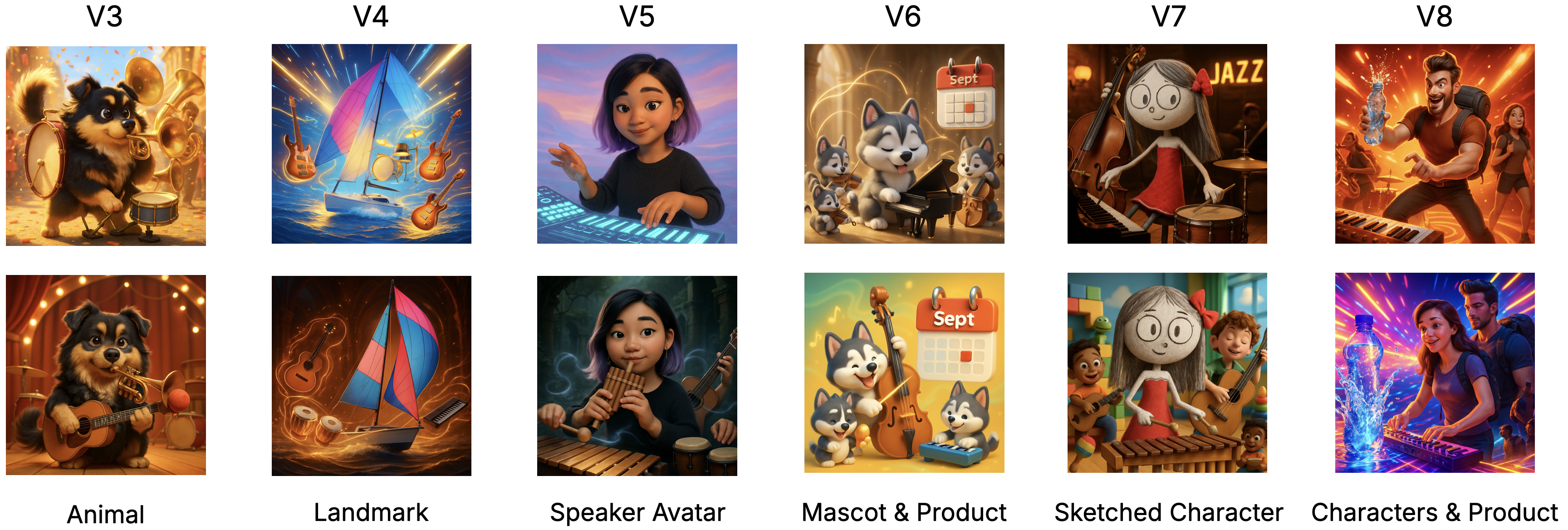}
  \caption{VidTune thumbnails grounded in participants’ own videos from the exploratory study. Diverse video anchors are used as thumbnail subjects.}
\label{fig:exploratory_thumbnails}
\Description{Two-row grid of 12 VidTune thumbnails labeled V3–V8 across six anchors—Animal, Landmark, Speaker Avatar, Mascot \& Product, Sketched Character, and Characters \& Product—showing: a dog drummer/guitarist on stage; a sailboat amid floating guitars/drums; a girl at a glowing synth and playing flute; a husky mascot band beside a calendar product; a sketchy pianist alone and with children; and a duo with a bottled-water product and keyboard in neon action lighting.}
\end{figure*}

We recruited 6 creators (2 professionals, 4 amateurs) through mailing lists and social media (P8, P11, P16, P20–P22). One participant (P8) had also joined the formative study, and three others (P11, P16, P20) took part in the controlled study. Consistent with the formative study, we included a deaf creator (P8) to understand whether VidTune’s visual approach could improve accessibility for soundtrack generation. All communication with P8 was text-based.
Before the study, we collected an edited video (without music) from each participant. 
Participants' videos ranged from 1-4 minutes and spanned diverse genres (Table~\ref{tab:exploratory_video_materials}). 3 out of 6 videos had narration (V8-V10).
Figure~\ref{fig:exploratory_thumbnails} illustrates resulting thumbnails grounded in participants’ videos, including speaker avatars, characters, animals, landmarks, and products.
During a 1hr study, we provided a tutorial of VidTune, invited participants to add music to their own video with VidTune, and asked semi-structured interview questions about their experience. We compensated \$40. 

\subsection{Findings}

\ipstart{Perceptions on reusing video content}
P20, who saw her product calendar appear alongside a mascot in the thumbnail, described the experience as more personal. During the study, she mainly focused on thumbnails instead of switching back and forth to the video player, explaining \textit{``I already know my video since I made it, and because the thumbnails include context, I can just review them instead of rewatching.''}
P8 was excited to see her dog featured in the thumbnail: \textit{``It's so cute, I want to share all of these [thumbnails] with my family and on my channel.''}
P16, who generated music for her daily vlog featuring herself, expressed mixed feelings. She noted, \textit{``It's cool that it [VidTune] got my purple hair and black knit correct. But some of the thumbnails don’t feel like me.''} While she still preferred stylized avatars over photorealistic generations, she explained that even minor mismatches could be distracting because she could easily spot differences in her own appearance.

\ipstart{Accessibility to diverse audiences}
P21 used VidTune to add music to his 5-year-old daughter’s sketch-style storytelling videos. 
He found it easier to generate child-appropriate music than searching stock libraries, with VidTune suggesting relevant keywords
After seeing the visual thumbnails, P16 was enthusiastic about how the sketch was animated into a character that performs instruments. He highlighted how VidTune could lower the barriers to music generation for children and novices: \textit{``Kids will love it, and parents will love it. There are more and more young creators now, and tools like this help them easily add music and learn music.''}

P8, a deaf creator, explained how thumbnails improve accessibility. \textit{``Very helpful. This [VidTune] lets us see the music. I used to only use tracks with lyrics as I can read them and understand, but now I can visualize.''} As she got closer to finalizing creation, she wanted more detailed information to confirm her choices—especially since the music would be shared publicly—and suggested that interactive question-answering could support this.
She also appreciated the fit check, which increased confidence in her selections, and noted that she wants to share VidTune with many other deaf creators who often struggle with adding music.

\ipstart{Compatibility with current workflows}
P11, who often reuses the same track across multiple videos, asked whether VidTune could adapt thumbnails to new contexts -- \textit{``If I use this music for another video about cars, will the thumbnails change to cars instead of sailboats? That would be cool, but I can also see it losing some of the `memory' I had.''} 
She also wanted to use trending songs, either to show them in VidTune or to generate similar tracks, explaining that while novel music is valuable, familiar references can sometimes help audiences connect more quickly.
P22, a professional filmmaker who frequently uses stock music, noted that VidTune-style thumbnails could help people skim large catalogs on stock sites. However, he emphasized that as a professional, he is accustomed to reviewing music with waveforms and therefore often prefers this standardized representation. 
Both P11 and P20 requested lightweight organization aligned with their current practice, such as renaming tracks and adding notes in their own words. 

\section{Discussion}


\subsection{Scope and Design Boundaries}


\revised{\textbf{Users.} VidTune supports a diverse range of users from novices to those with limited hearing, who can benefit from prompt suggestions, contextual thumbnails, and fit-evaluation tools. 
In our studies, more experienced creators also found the thumbnails useful for rapid skimming and communication, but noted that they would still rely on existing tools for fine-grained control, so we view VidTune as a complementary exploration aid rather than a full expert workstation. 
Although we did not study children directly, some participants suggested that the thumbnails could help users with shorter attention spans such as children.}

\textbf{Musical Content.}  VidTune targets non-lyrical (instrumental) music. Extending to vocal tracks would introduce additional review needs (\textit{e.g.,} lyric semantics, vocal style, timing, and prosody) and require multimodal support beyond images, such as highlighting key lyrics synchronized to beats (\textit{e.g.,} karaoke-style phrasing).

\textbf{Control mechanisms.} VidTune currently accepts text prompts and does not support reference audio. It also omits structural scaffolding before generation (\textit{e.g.,} outlining beats/chords~\cite{huang2016chordripple}), fine-grained mixing/mastering~\cite{clemens2025mixassist}, and automatic video editing to fit music~\cite{zhang2025let}. 
\revised{Our formative study and prior work~\cite{hammad2025s} show that editors often try to match music beats to specific cuts, but precise beat-sync requires either retiming the video (adjusting or inserting cuts) or fine-grained temporal control over generated music. Since current text-to-music models mainly allow control over global attributes such as mood and overall tempo rather than beat-level changes, We focus on helping creators select and place soundtracks at the scene level, leaving fine-grained alignment to downstream editing tools.
Future work could explore how to incorporate model advances in parameterized controls~\cite{musicControlNet} and beat/phrase/stem extraction while maintaining simple, intuitive interactions.}

\subsection{Challenges in Generative Abundance} 
By producing many candidates in a short time with minimal user input, generative models shift creative effort from making to choosing. This shift introduces new challenges. 
First, homogenization can emerge~\cite{anderson2024homogenization, han2025poet, 10.1145/3635636.3656204}, especially for novices who are unaware of the broader space or how to reach it.
We therefore set diversity as a design goal (D2) and built VidTune to expand users' prompts and provide diverse music generations.
Second, with abundant choices, review and comparison become the bottleneck, both in terms of time and cognitive load. 
Building on sensemaking work across text~\cite{reza2024abscribe, gero2024supporting}, image~\cite{huh2023genassist, almeda2024prompting}, and video~\cite{huh2025videodiff}, VidTune makes generative music glanceable via contextual thumbnails.

Beyond reviewing burden, a selection-centric workflow could dilute agency and ownership~\cite{yuan2022wordcraft, huh2025videodiff} as users may feel less attached to an artifact they spent less time on.
Future tools can mitigate this by adding intentional (good) friction\revised{~\cite{chen2024exploring, cox2016design}} that elicits meaningful user input before assistance and by personalizing suggestions to the creator’s evolving tastes and project context. As GenAI reshapes creative practices, interfaces should support creators to explore broadly, compare efficiently, and feel ownership over the result.

\subsection{Making Music More Visual}
VidTune is grounded in the idea of making music more visual -- translating soundtracks into forms that can be skimmed, compared, and remembered through sight, drawing on the speed and parallelism of visual perception. 
Our study showed that thumbnails capturing key musical qualities helped users understand the music \cite{10.1145/3461835}, filter irrelevant tracks, aid recall \cite{childers1984conditions}, and—when contextualized with their video—foster a more personal and enjoyable experience. That said, visualizing music carries inherent limitations. Proxies may not fully capture the richness of sound, mappings between musical and visual features could vary across cultures and individuals \cite{walker1987effects}, and users may risk privileging what \textit{``looks right''} over what truly sounds right. It is therefore crucial to frame such visuals as augmentations to listening, not replacements for it. 

While we used thumbnails primarily for sensemaking in soundtrack creation, participants were excited about use cases beyond, including richer music listening with thumbnails or repurposing them for music videos or album covers. Another potential is ~\textit{music-driven storytelling}, where thumbnails expand into storyboards to spark narrative ideas from music. Finally, future work could also explore music editing through visual manipulations of thumbnails, \textit{e.g.,} enlarging a cello to amplify its volume or substituting it with a double bass to alter orchestration.

\subsection{Building with Imperfect Gen AI Models}
VidTune relies on LMMs for scene analysis and prompt suggestions, a text-to-music model for generating soundtracks, and image and video generation models for generating thumbnails. 
These models have limitations and are prone to errors. For instance, suggesting music prompts based on the scene can reflect bias in how LMMs interpret visual context~\cite{BuolamwiniG18}. 
Music generators may miss user intent for culturally specific styles or underrepresented genres -- (\textit{e.g.,} collapsing diverse African traditions into generic ``tribal drums''), echoing longstanding concerns about dataset and modeling bias in language models~\cite{Bender2021-BENOTD-6} and image models~\cite{luccioni2023stablebiasanalyzingsocietal}.
Finally, biases reported in image generation can also surface when associating musical qualities with visuals in thumbnail generation. For instance, when multiple characters are involved, mappings can drift toward stereotypical assignment 
(\textit{e.g.,} male characters for ``energetic'', female for ``calm'')~\cite{han2025poet, luccioni2023stablebiasanalyzingsocietal}.
To work around and address these issues, future tools should offer transparent, user-facing explanations of how recommendations are generated. They can also intentionally diversify options when uncertainty is high, and provide a feedback loop for users to flag when outputs do not align with requests.
\subsection{Role of Playfulness in Creative Tasks}
While we did not set out to design VidTune as a playful system, participants reported unexpected enjoyment: reusing their own characters and scenes in thumbnails made reviewing feel personal, and the visuals reduced the tedium of sequential listening. Prior work on gamification shows that making progress visual can sustain attention and momentum~\cite{hamari2014does}; VidTune's thumbnails provide that visibility.
As generative AI systems increasingly automate production and shift effort toward intent articulation and review, we argue that creativity tools should preserve the ~\textit{fun} parts with the creators. 
In practice, this means offloading tedious steps while making input and review more engaging. VidTune addresses the review side, while complementary works have explored playful input (\textit{e.g.,} toy-play interactions for storytelling with LLMs~\cite{chung2024toyteller}). We believe future creativity tools should cultivate enjoyable micro-moments to make the process as rewarding as the outcome.

\section{Conclusion}
We introduced \textit{VidTune}, a system that helps video creators generate soundtracks by expanding prompts, visualizing music with video-grounded thumbnails, and supporting iterative refinement. By making music more \emph{visible in context}, VidTune shifts the effort from sequential listening to rapid, informed comparison.
\revised{Through a technical evaluation, we showed that VidTune’s contextual thumbnails more faithfully reflected musical attributes than baseline thumbnails. In a user study (N=12) comparing VidTune to a baseline text-to-music interface, participants reported that VidTune made it easier to understand, compare, and remember music candidates, and described the workflow as more expressive and enjoyable.
Finally, a case study (N=6) with creators' own videos showed that participants valued seeing their footage reflected in the thumbnails and felt this made the generated music feel more tailored to their content.}
Broadly, this work demonstrates how generative AI shifts soundtrack selection for video from a retrieval task to a creative endeavor, with VidTune illustrating how music creation can be more visual, playful, and accessible for diverse creators.





\begin{acks}
We thank Nick Bryan for his valuable feedback, and Gabi Duncombe for generously allowing us to use her videos in this work. Mina Huh is supported by a Google Ph.D. fellowship.
\end{acks}


\bibliographystyle{ACM-Reference-Format}
\bibliography{sample-base}

\clearpage
\onecolumn
\appendix

\section{PARTICIPANTS}

\begin{table*}[htbp!]
\small\sffamily\def\arraystretch{1}\setlength{\tabcolsep}{0.8em}
    \centering
    \begin{tabular}{lllllll}
        \toprule
       PID  & Gender & Age & Job & Videos Created & AI Music Experience & Hearing\\
       \midrule
        1 & Male & 43 & Magician & Demonstrations, AI films & Regular & Hearing \\ 
        2 & Female & 42 & AI filmmaker & AI films, Documentaries & Regular & Hearing \\
        3 & Male & 34 & Illustrator & Motion graphics, commercials & Never & Hearing \\
        4 & Female & 26 & Motion designer & Reels, Sports highlights & Occasional & Hearing \\
        5 & Female & 45 & ASL teacher & Teaching materials & Never & Deaf (Profound) \\
        6 & Male & 46 & Design instructor & Student testimonials & Regular & Hearing \\
        7 & Female & 36 & Podcast producer & Podcasts & Never & Deaf (Moderate) \\
        8 & Female & 35 & ASL teacher & Teaching materials, Travel vlogs  & Never & Deaf (Profound) \\
        9 & Female & 22 & Technical manager & Vlogs & Never & Hearing \\
        10 & Female & 24 & Product manager & Vlogs, Short-forms & Never & Hearing \\
        11 & Female & 45 & Content creator & Sports recaps, How-To videos & Never & Hearing \\
        12 & Female & 26 & Software engineer & Skits, Weddings & Never & Hearing \\
        13 & Male & 27 & Account manager & Music videos, AI films & Regular & Hearing \\
        14 & Female & 27 & Filmmaker & Music videos, Animated films & Regular & Hearing \\
        15 & Male & 23 & Technology analyst & Vlogs, Short-forms & Occasional & Hearing \\
        16 & Female & 30 & Content creator & Vlogs, UX Tutorials & Regular & Hearing \\
        17 & Female & 30 & Campaign manager & Short-forms & Never & Hearing \\
        18 & Male & 45 & Designer & Explainers, Travel Vlogs & Occasional & Hearing \\
        19 & Female & 33 & Software engineer & Music performance videos, Vlogs & Never & Hearing \\
        20 & Female & 23 & Product manager & Educational/lecture videos & Occasional & Hearing \\
        
        21 & Male & 38 & Freelance video editor & Animations, Family videos & Occasional & Hearing \\
        22 & Male & 56 & Filmmaker & Animations, AI films, Ads & Regular & Hearing \\
        \bottomrule
    \end{tabular}
    \caption{Demographics of study participants (Formative study (N=8): P1-P8; Evaluation study (N=12): P9-P20; Exploratory study (N=6): P8, P11, P16, P20-P22)} 
    \label{tab:participants}
    \Description{Demographics table of 22 participants (14 female, 8 male; ages 22–56) with columns for PID, job (e.g., filmmaker, content creator, ASL teacher), videos created, AI-music experience (Regular/Never/Occasional), and hearing status—mostly hearing with 3 Deaf (2 profound, 1 moderate)—covering cohorts from the formative (P1–P8), evaluation (P9–P20), and exploratory (P8, P11, P16, P20–P22) studies.
}
\end{table*}

\section{VIDEO MATERIALS}

\begin{table}[htbp!]
\sffamily\def\arraystretch{1.05}\setlength{\tabcolsep}{0.6em}
\small
  \centering
  \begin{tabular}{ccccc}
    \toprule
    \textbf{Video ID} & \textbf{Video Type} & \textbf{Duration} & \textbf{Thumbnail Anchor} & \textbf{Studies used} \\ \midrule
    V1 & AI-generated film         & 0:32 & Animated Characters                   & \S\ref{sec:technical_eval}, \S\ref{sec:user_eval} \\ 
    V2 & AI-generated film    & 0:44 & Animated Characters    & \S\ref{sec:technical_eval}, \S\ref{sec:user_eval} \\ 
    V3 & AI-generated film   & 0:42 & Animated Humans          & \S\ref{sec:technical_eval} \\ 
    V4 & AI-generated film   & 2:00 & Animated Humans          & \S\ref{sec:technical_eval} \\ 
    V5 & Vlog (CC-licensed)~\cite{ho_paris_2025_youtube}     & 15:03 & Landmark       & \S\ref{sec:technical_eval} \\ 
    V6 & AI-generated film       & 0:58 & Animated Characters       & \S\ref{sec:technical_eval}, \S\ref{sec:user_eval} \\ 
    \bottomrule
  \end{tabular}
  \caption{Videos used in the technical evaluation (Section~\ref{sec:technical_eval}) and controlled user evaluation (Section~\ref{sec:user_eval}). All generated films were provided by an independent filmmaker and are used with consent. Participants' own videos used in the exploratory case study are in Table~\ref{tab:exploratory_video_materials}.}
  \label{tab:video_materials}
  \Description{Video materials table: six clips—V1 (0:32) and V2 (0:44) AI-generated films with Animated Characters used in §§6–7; V3 (0:42) and V4 (2:00) AI-generated films with Animated Humans used in §6; V5 vlog (CC, 15:03) with Landmark anchor used in §6; V6 (0:58) AI-generated film with Animated Characters used in §§6–7.}
\end{table}






\clearpage
\onecolumn
\appendix

\section{Pipeline Prompts}\label{sec:pipeline_prompts}

\begin{table*}[!h]
\small
\resizebox{0.9\textwidth}{!}{%
\begin{tabular}{@{}p{15cm}@{}}
\toprule
\textbf{1. Video Analysis \& Prompt Suggestions} \\ \midrule
\textbf{Task}: Analyze uploaded videos to extract scenes with timestamps and music guidance \\[4pt]
\textbf{Prompt Template}:\\
Analyze this video and segment it into meaningful scenes. For each scene, provide detailed music composition guidance.\\[2pt]
Please return a JSON response with this EXACT structure:\\[2pt]
\{\\
\quad ``videoAnalysis'': \{\\
\qquad ``totalDuration'': ``MM:SS'',\\
\qquad ``scenes'': [\\
\qquad\quad \{\\
\qquad\qquad ``sceneId'': 1,\\
\qquad\qquad ``startTime'': ``MM:SS'',\\
\qquad\qquad ``endTime'': ``MM:SS'',\\
\qquad\qquad ``duration'': ``MM:SS'',\\
\qquad\qquad ``sceneDescription'': ``Detailed description of what's happening visually'',\\
\qquad\qquad ``vibeDescription'': ``Emotional atmosphere, mood, and feeling of the scene'',\\
\qquad\qquad ``musicKeywords'': \{\\
\qquad\qquad\quad ``genres'': [``array of 10+ music and video genres that would fit this scene''],\\
\qquad\qquad\quad ``instruments'': [``array of 10+ instruments that would enhance this scene''],\\
\qquad\qquad\quad ``moods'': [``array of 10+ emotional/mood descriptors for music''],\\
\qquad\qquad\quad ``energy'': [``array of 10+ energy level and tempo descriptors'']\\
\qquad\qquad \}\\
\qquad\quad \}\\
\qquad ],\\
\qquad ``overallKeywords'': \{\\
\qquad\quad ``dominantGenres'': [``top 4 genres for the entire video''],\\
\qquad\quad ``primaryMoods'': [``top 4 moods for the entire video''],\\
\qquad\quad ``recommendedInstruments'': [``top 4 instruments for the entire video''],\\
\qquad\quad ``energyProfile'': ``overall energy description'',\\
\qquad\quad ``tempoRange'': ``recommended tempo range''\\
\qquad \}\\
\quad \}\\
\}\\[4pt]
Requirements:\\
1.\ Segment video into 2--6 logical scenes based on visual/narrative changes\\
2.\ Each scene should be 15--60 seconds long\\
3.\ Provide 10+ options for each music keyword category per scene\\
4.\ Focus on music production keywords that would help generate appropriate background music\\
5.\ Consider cinematic, documentary, and video content music styles\\
6.\ For genres: include both music genres (electronic, orchestral, acoustic) and video/cinematic genres (thriller, romance, action)\\
7.\ For energy: include both energy levels (high, medium, low) and tempo descriptors (fast, slow, building, driving)\\
8.\ Ensure keywords are diverse and cover different musical approaches for each scene\\[2pt]
Video duration: \{videoDuration\} seconds\\[2pt]
Return ONLY valid JSON, no additional text or explanation.\\[8pt]

\textbf{Template Variables}: \{videoDuration\} \\
\bottomrule
\end{tabular}
}
\caption{Video analysis prompt.}
\label{tab:video_analysis}
\end{table*}

\begin{table*}[!h]
\small
\resizebox{0.9\textwidth}{!}{%
\begin{tabular}{@{}p{15cm}@{}}
\toprule
\textbf{2. Music Analysis: Fit Check \& Tags (Part 1)} \\ \midrule
\textbf{Task}: Analyze generated music tracks using audio input for fit evaluation and tagging \\[4pt]
\textbf{Prompt Template}:\\
You are a music expert with the ability to listen to and analyze audio. Listen carefully to this music track and provide comprehensive analysis based on what you actually hear in the audio.\\[2pt]
Base your analysis only on what you hear in the attached audio file. Ignore any unrelated text context.\\
You must identify and include all instruments and sounds you hear in the audio.\\[6pt]

Reference info (for final fit evaluation only):\\
-- Original User Prompt: ``\{originalPrompt\}''\\
-- User Video Metadata: title ``\{title\}'', type ``\{videoType\}'', audience ``\{audience\}'', goal ``\{soundtrackGoal\}''\\
-- Video Scene Data (current reference): description ``\{sceneDescription\}'', vibe ``\{sceneVibe\}'', duration ``\{sceneDuration\}'' (seconds); start ``\{sceneStart\}''--end ``\{sceneEnd\}'' (seconds)\\[6pt]

Analysis task:\\
Listen to the attached audio file and analyze what you hear:\\[2pt]
1.\ Fit evaluation: Evaluate how well it fits the original user prompt and the current scene/video context.\\
\quad -- Good fit: provide 2--3 specific reasons (4--7 words each) why the music works well\\
\quad -- Bad fit: provide 1--2 specific reasons (4--7 words each) why the music does not work well\\[2pt]
Examples of good responses:\\
\quad -- ``Acoustic guitar fits family content mood''\\
\quad -- ``Medium tempo aligns with scene energy''\\
\quad -- ``Genre matches dominant video themes''\\[2pt]
Examples of bad responses:\\
\quad -- ``Tempo too fast for slow scenes''\\
\quad -- ``Genre mismatch with video content''\\
\quad -- ``Energy level too high for calm scenes''\\[2pt]
Avoid vague statements like ``doesn't match'', ``unspecified genre'', ``not suitable''.\\[6pt]

2.\ Prominent tags: Generate 3 tags based on what you hear in the audio.\\
\quad -- 1 genre tag (Purple: \#8B5CF6)\\
\quad -- 1 mood/emotion tag (Green: \#10B981)\\
\quad -- 1 instrument/technical tag (Blue: \#06B6D4)\\[2pt]
Instrument tag requirement: choose the most audible/prominent instrument (drums, piano, guitar, bass, strings, brass, synthesizer, etc.).\\
\bottomrule
\end{tabular}
}
\caption{Music analysis prompt (part 1).}
\label{tab:gemini_audio_analysis_part1}
\end{table*}

\begin{table*}[!h]
\small
\resizebox{0.9\textwidth}{!}{%
\begin{tabular}{@{}p{15cm}@{}}
\toprule
\textbf{3. Music Analysis: Image Description \& Details (Part 2)} \\ \midrule
3.\ Image description: Create a complete visual description based on what you hear in the music.\\
-- Include all instruments you detect\\
-- Show the protagonist character interacting with all instruments\\
-- Match visual energy to musical energy\\
-- Refer to the character as ``the protagonist character''\\
-- Use 4--5 sentences covering all instruments and elements\\[6pt]

\textbf{Image prompt generation instructions}:\\
For each track, create a detailed image prompt using this structure:\\[4pt]
\textbf{Base character}: ``\$\{protagonist\}''\\[4pt]
Then add these elements based on the music analysis (examples are guidance, not fixed choices):\\[6pt]

1.\ Visual scene \& genre (examples):\\
\quad -- Jazz: ``A warm, intimate jazz club with soft amber lighting and swirling smoke patterns''\\
\quad -- Electronic: ``A futuristic urban landscape under a neon glow, with energy conduits crisscrossing through digital architecture''\\
\quad -- Orchestral: ``A grand natural amphitheater surrounded by majestic mountains and flowing waterfalls''\\
\quad -- Rock: ``A dynamic concert stage with dramatic lighting and electric energy crackling through the air''\\
\quad -- Ambient: ``A serene, ethereal space with floating geometric forms and gentle light emanations''\\
\quad -- Folk: ``A cozy, rustic environment with warm wooden textures and natural, earthy elements''\\[6pt]

2.\ Protagonist performance (examples):\\
\quad -- \$\{videoType === 'octopus' ? 'Her tentacles gracefully embody the musical instruments' : 'She gracefully embodies the musical instruments'\} -- [describe specific instruments from the caption]\\
\quad -- ``The visual prominence and size of each instrument effect correlates with its prominence in the music''\\[6pt]

3.\ Implied motion \& tempo (examples):\\
\quad -- Fast/energetic: ``rapid, energetic movements with sharp light trails, motion blur, and dynamic speed lines''\\
\quad -- Slow/calm: ``flowing, graceful movements with smooth light trails and gentle undulating patterns''\\
\quad -- Medium: ``rhythmic, measured movements with steady light pulses and balanced visual flow''\\[6pt]

4.\ Emotion (examples):\\
\quad -- Happy/joyful: ``joyful expression with bright, sparkling eyes and an uplifted, confident posture''\\
\quad -- Calm/peaceful: ``serene expression with peaceful, steady eyes and relaxed, flowing posture''\\
\quad -- Energetic: ``enthusiastic expression with intense, focused eyes and dynamic, powerful gestures''\\
\quad -- Mysterious: ``enigmatic expression with knowing, slightly narrowed eyes and graceful, controlled movements''\\[6pt]

5.\ Color palette: Follow overall trends from color theory; use harmony or contrast to reinforce mood.\\
\quad -- Warm, high-energy palettes often leverage complementary contrasts (reds/oranges against blues) to emphasize vibrancy.\\
\quad -- Calm or peaceful palettes favor analogous, low-saturation hues for balance and unity.\\
\quad -- Dark or mysterious palettes rely on deep values and limited contrast to build intrigue.\\
Examples:\\
\quad -- Positive/energetic: ``vibrant, warm colors dominated by golden yellows, energetic oranges, and uplifting blues''\\
\quad -- Calm/peaceful: ``soft, pastel tones with gentle blues, warm creams, and subtle lavenders''\\
\quad -- Mysterious/dark: ``dark, rich colors with deep purples, midnight blues, and ethereal silver accents''\\[6pt]

End with: ``Style: 3D animation, Pixar-quality rendering, cinematic lighting, highly detailed, vibrant and expressive.''\\[8pt]

Return only a JSON object with this format:\\[2pt]
\{ ``fitAnalysis'': \{...\}, ``prominentTags'': [...], ``imageDescription'': ``...'', ``detailedMusicDescription'': ``...'' \}\\[6pt]

Detailed music description must cover tempo, time signature, key, instrumentation, form, genre specificity, articulation, production, emotional character, and distinctive elements. The description should be comprehensive enough to accurately recreate or edit the musical content.\\[8pt]

\textbf{Template Variables}: \{originalPrompt\}, \{trackTitle\}, \{fullQuery\}, \{videoMetadata\}, \{sceneData\} \\
\bottomrule
\end{tabular}
}
\caption{Music analysis prompt (part 2).}
\label{tab:gemini_audio_analysis_part2}
\end{table*}

\begin{table*}[!h]
\small
\resizebox{0.9\textwidth}{!}{%
\begin{tabular}{@{}p{15cm}@{}}
\toprule
\textbf{4. Edit Request Expansion} \\ \midrule
\textbf{Task}: Expand a user's edit request into 4 creative variations that minimally modify the original while reflecting the user's intent. \\[4pt]

\textbf{Prompt Template}:\\
You are a music production expert. A user wants to edit an existing music track with this request: ``\{editRequest\}''\\[2pt]
Original music description: ``\{originalMusicDescription\}''\\
Original prompt: ``\{originalPrompt\}''\\[6pt]

Critical requirements:\\
1.\ Intent-focused editing: ensure the user's intent is clearly reflected throughout the description. If they say ``make it calmer'', the overall feel should be noticeably calmer.\\
2.\ Strategic changes: modify all relevant elements that support the user's intent (tempo, energy, instruments, dynamics, mood).\\
3.\ Preserve core identity: keep the genre and main instrumentation unless specifically requested to change; adapt energy/mood/tempo as needed.\\[6pt]

Your task: create 4 conservative variations that minimally modify the original description.\\[6pt]

Examples of intent-focused editing (illustrative, not fixed choices):\\
Original: ``Upbeat electronic dance track with pounding drums and bright synths''\\
User request: ``make it calmer''\\
-- Variation 1 Title: ``Calmed Down'' (description: ``Relaxed electronic ambient track with soft drums and mellow synths'')\\
-- Variation 2 Title: ``Reduced Energy'' (description: ``Gentle electronic chillout track with subtle drums and warm synths'')\\
-- Variation 3 Title: ``Slower Pace'' (description: ``Laid-back electronic downtempo track with gentle drums and soothing synths'')\\
-- Variation 4 Title: ``Softened'' (description: ``Peaceful electronic meditation track with quiet drums and ethereal synths'')\\[6pt]

Title requirements (must follow exactly):\\
-- Maximum 2--4 words\\
-- Simple verbs: Added, Removed, Slower, Faster, Softer, Louder, Calmer, Energetic\\
-- No abstract terms: avoid ``overall'', ``rhythmic dive'', ``dynamic'', ``enhanced'', ``textural''\\[4pt]

Title examples (copy these patterns):\\
-- User says ``add piano'' $\rightarrow$ ``Added Piano''\\
-- User says ``make slower'' $\rightarrow$ ``Slower Tempo''\\
-- User says ``more energy'' $\rightarrow$ ``More Energy''\\
-- User says ``remove drums'' $\rightarrow$ ``No Drums''\\
-- User says ``calmer'' $\rightarrow$ ``Calmed Down''\\
-- User says ``louder'' $\rightarrow$ ``Louder Volume''\\[6pt]

Return only a JSON response with this exact format:\\[2pt]
\{\\
\quad ``variations'': [\\
\qquad \{ ``description'': ``Original description with targeted changes throughout to fulfill user intent'', ``title'': ``MAXIMUM 2--4 WORDS describing the change'', ``emphasis'': ``What specifically was changed'' \},\\
\qquad \{ ``description'': ``Original description with targeted changes throughout to fulfill user intent'', ``title'': ``MAXIMUM 2--4 WORDS describing the change'', ``emphasis'': ``What specifically was changed'' \},\\
\qquad \{ ``description'': ``Original description with minimal targeted changes'', ``title'': ``Just the edit change (2--3 words max)'', ``emphasis'': ``What specifically was changed'' \},\\
\qquad \{ ``description'': ``Original description with minimal targeted changes'', ``title'': ``Just the edit change (2--3 words max)'', ``emphasis'': ``What specifically was changed'' \}\\
\quad ]\\
\}\\[8pt]

\textbf{Template Variables}: \{editRequest\}, \{originalMusicDescription\}, \{originalPrompt\} \\
\bottomrule
\end{tabular}
}
\caption{Edit request expansion prompt.}
\label{tab:edit_request_expansion}
\end{table*}

\begin{table*}[!h]
\small
\resizebox{0.9\textwidth}{!}{%
\begin{tabular}{@{}p{15cm}@{}}
\toprule
\textbf{5. Blend-Based Expansion} \\ \midrule
\textbf{Task}: Given two or more music descriptions, infer their common attributes, produce one detailed common description, and suggest 4 short variations that remain similar to the given examples. \\[4pt]

\textbf{Prompt Template}:\\
You are a music production expert. Identify common attributes across multiple reference music descriptions and propose closely related variations.\\[4pt]
Input music descriptions (2 or more):\\
\{\\
\quad ``desc\_1'': ``\{musicDescriptions[0]\}'',\\
\quad ``desc\_2'': ``\{musicDescriptions[1]\}'',\\
\quad ``desc\_N'': ``\{musicDescriptions[N]\}''\\
\}\\[6pt]

Analysis goal: derive a unified, detailed common description that captures overlapping style, genre/sub-genre, tempo/energy, instrumentation roles, rhythmic feel, harmonic palette, production aesthetics (space/reverb/stereo image/dynamics), and emotional character that are shared by the inputs. Preserve elements that consistently recur; avoid outliers unique to only one example.\\[6pt]

Variation goal: propose four short postfix-style variations (2--4 words each) that keep the music similar to the references while exploring small, controlled differences (tempo nudges, instrument emphasis shifts, articulation tweaks, mix emphasis). Titles must follow the same rules as in edit expansion: short, literal, non-abstract.\\[6pt]

Examples (illustrative, not fixed choices):\\
References suggest: ``Organic lo-fi chill with soft drums, warm electric piano, mellow bass, gentle sidechain, relaxed 80--90 BPM, intimate stereo, cozy mood.''\\
-- Variation: ``Added Piano'' (emphasis: ``bring out warm e-piano voicings'')\\
-- Variation: ``Softer Drums'' (emphasis: ``brush/loose kit, reduced transients'')\\
-- Variation: ``Deeper Bass'' (emphasis: ``rounder low end, sustained notes'')\\
-- Variation: ``Airy Reverb'' (emphasis: ``slightly longer tails, wider space'')\\[6pt]

Return only a JSON object with this exact format:\\[2pt]
\{\\
\quad ``commonDescription'': ``Detailed unified description capturing shared attributes (genre/sub-genre, tempo/energy, rhythmic feel, instrumentation roles, harmonic tendencies, production aesthetics, emotional character).'',\\
\quad ``variations'': [\\
\qquad \{ ``title'': ``2--4 words'', ``emphasis'': ``what to tweak while staying similar'' \},\\
\qquad \{ ``title'': ``2--4 words'', ``emphasis'': ``what to tweak while staying similar'' \},\\
\qquad \{ ``title'': ``2--4 words'', ``emphasis'': ``what to tweak while staying similar'' \},\\
\qquad \{ ``title'': ``2--4 words'', ``emphasis'': ``what to tweak while staying similar'' \}\\
\quad ]\\
\}\\[8pt]

\textbf{Template Variables}: \{musicDescriptions\} \\
\bottomrule
\end{tabular}
}
\caption{Blend-based expansion prompt.}
\label{tab:blend_request_expansion}
\end{table*}


\end{document}